\pdfoutput=1

\documentclass[prd,aps,amsmath,nofootinbib,amssymb,eqsecnum,showkeys,twocolumn]{revtex4-2}

\usepackage{amsmath, latexsym, amssymb, graphicx, color, multirow, bm, braket, ascmac, bbm, outlines, amsfonts}
\usepackage{hyperref}
\usepackage{mathtools}
\usepackage{subcaption}
\usepackage{lineno}
\usepackage{cancel}
\usepackage{dsfont}
\usepackage[table]{xcolor}
\usepackage{tabularx}
\usepackage[T1]{fontenc}
\usepackage[capitalize]{cleveref}
\usepackage{romannum}
\usepackage{ragged2e}
\usepackage{makecell, physics}
\definecolor{nicered}{rgb}{.7,.1,.1}
\definecolor{nicegreen}{rgb}{.1,.5,.1}
\definecolor{darkblue}{rgb}{0,0,.55}
\definecolor{burntorange}{rgb}{0.8, 0.33, 0.0}
\definecolor{bluee}{rgb}{0,0.1,0.9}
\definecolor{crimson}{rgb}{0.86, 0.08, 0.24}
\hypersetup{colorlinks, citecolor=nicegreen,linkcolor=nicered, urlcolor=darkblue}
\usepackage{multirow}
\usepackage{slashed}
\usepackage{comment}
\usepackage{ulem}
\numberwithin{equation}{section}
\usepackage{verbatim}
\newcommand{\RNum}[1]{\uppercase\expandafter{\romannumeral #1\relax}}

\newcommand\mrm[1]{{\color{magenta}#1}}

\begin{document}
\captionsetup{justification=RaggedRight}
\pagenumbering{arabic}
\hfill ACFI T23-03
	\title{Testing Complex Singlet Scalar Cosmology at the Large Hadron Collider
		\\
	}
	\author{Wenxing Zhang$^{1}$, Yizhou Cai$^{2}$, Michael J. Ramsey-Musolf$^{1,3}$, Lei Zhang$^{2}$ }
	\affiliation{\vspace{2mm} \\
		$^1$Tsung-Dao Lee Institute and School of Physics and Astronomy, \\Shanghai
		Jiao Tong University, 800 Dongchuan Road, Shanghai 200240, China \\
  		$^2$ Department of Physics, Nanjing University, 22 Hankou Road, Nanjing 210093, China \\
		$^3$Amherst Center for Fundamental Interactions, Department of Physics, University of Massachusetts, Amherst, MA 01003, USA \\}

    \begin{abstract}
	
The Standard Model extended with a complex singlet scalar (cxSM) can admit a strong first order electroweak phase transition (SFOEWPT) as needed for electroweak baryogenesis and provide a dark matter (DM) candidate.
The presence of both a DM candidate and a singlet-like scalar that mixes with the Standard Model Higgs boson leads to the possibility of a  $b\bar{b}+\text{MET}$ final state in $pp$ collisions. Focusing on this channel, 
we analyze the prospective reach at the Large Hadron Collider (LHC) for a heavy singlet-like scalar in regions of cxSM parameter space compatible with a SFOEWPT and DM phenomenology.
We identify this parameter space while implementing current constraints from electroweak precision observable and Higgs boson property measurements as well as those implied by LHC heavy resonance searches.
Implementing a proposed search strategy, we find that the heavy scalar and DM candidate can be probed up to 1 TeV and 450 GeV at 2$\sigma$ level respectively.

\end{abstract}
	\maketitle
	
    \section{Introduction}

The origin of the cosmic baryon asymmetry is one of the long-standing puzzles in particle physics. Electroweak baryogenesis ~\cite{Rubakov:1996vz, Funakubo:1996dw, Quiros:1999jp, Morrissey:2012db} provides a promising solution and can be tested at the current collider experiments\cite{Ramsey-Musolf:2019lsf,Morrissey:2012db}.
In general, a baryogenesis mechanism should meet Sakharov's three conditions~\cite{Sakharov:1967dj, Kajantie:1995kf}:

\begin{itemize}
	\item  
	Baryon number violating interactions.
	
	\item 
	C and CP violation.
	
	\item 
	Departure from thermal equilibrium (or CPT violation).
	
\end{itemize}

The baryon number violating processes could appear in Standard Model (SM)  via the non-perturbative effects caused by sphaleron transitions ~\cite{Klinkhamer:1984di, McLerran:1989ab}. 
In principle, the requisite CP violation also appears in SM via Cabibbo-Kobayashi-Maskawa matrix, though the strength is found to be insufficient to generate the observed matter-antimatter asymmetry.

In the SM, a possible departure from thermodynamic equilibrium could happen via a first order electroweak phase transition (FOEWPT) at the electroweak temperature, $T_\mathrm{EW}\sim 140$ GeV, that marks the onset of electroweak symmetry-breaking\cite{Ramsey-Musolf:2019lsf}. To ensure preservation of any baryon asymmetry produced during this transition, the latter must be sufficiently strong. 
The occurrence of a FOEWPT requires the mass of the Higgs boson to lie below $\sim$ 70 GeV~\cite{Csikor:1998eu, Bochkarev:1987wf, Kajantie:1996qd, Kajantie:1995kf, Kajantie:1996mn, Delaunay:2007wb, Gould:2019qek, Andersen:2017ika, DOnofrio:2015gop}, which is inconsistent with the experimental observation~\cite{ATLAS:2012yve, CMS:2012qbp}.

Therefore,  electroweak baryogenesis can only be realised in extensions of SM that accommodate a strongly first order electroweak phase transition (SFOEWPT).  
The most widely considered scenarios include
the real singlet extensions (xSM)~\cite{Choi:1993cv, Ham:2004cf, Profumo:2007wc, Espinosa:2011ax, Profumo:2014opa, Damgaard:2015con, Espinosa:1993bs, Benson:1993qx, Vergara:1996ub, Noble:2007kk, Espinosa:2008kw, Barger:2007im, Ashoorioon:2009nf, Das:2009ue, Espinosa:2011ax, Cline:2012hg, Chung:2012vg, Huang:2012wn, Damgaard:2013kva, Fairbairn:2013uta, No:2013wsa, Profumo:2014opa, Craig:2014lda, Curtin:2014jma, Chen:2014ask, Katz:2014bha, Kozaczuk:2015owa, Kanemura:2015fra, Damgaard:2015con, Huang:2015tdv, Kanemura:2016lkz, Kotwal:2016tex, Brauner:2016fla, Huang:2017jws, Chen:2017qcz, Beniwal:2017eik, Cline:2017qpe, Kurup:2017dzf, Alves:2018jsw, Li:2019tfd, Gould:2019qek, Kozaczuk:2019pet, Carena:2019une, Heinemann:2019trx, Branco:1998yk, Zhang:2023jvh, Azatov:2022tii}, 
complex singlet extensions (cxSM) ~\cite{ Barger:2008jx, Chao:2014ina, Chiang:2017nmu, Gross:2017dan, Cheng:2018ajh, Grzadkowski:2018nbc, Chen:2019ebq, Cho:2021itv, Cho:2022zfg, Schicho:2022wty, Cho:2021itv}, 
Higgs doublet extensions~\cite{Land:1992sm, Cline:1996mga, Fromme:2006cm, Dorsch:2013wja, Hammerschmitt:1994fn, Turok:1991uc, Davies:1994id, Fromme:2006cm, Cline:2011mm, Dorsch:2013wja, Dorsch:2014qja, Andersen:2017ika, Harman:2015gif, Dorsch:2017nza, Bernon:2017jgv, Basler:2016obg}, 
and supersymmetric extensions~\cite{Pietroni:1992in, Espinosa:1993yi, Brignole:1993wv, Davies:1996qn, Funakubo:2002yb, Ham:2004nv, Cohen:2012zza, Carena:1996wj, Delepine:1996vn, Cline:1996cr, Laine:1998qk, Carena:2008vj, Laine:2012jy, Curtin:2012aa, Carena:2012np, Katz:2015uja}. 

Among the distinctive signatures of such mixing is resonant di-Higgs production, where the heavy resonance is a mixed singlet-doublet state~\cite{Ramsey-Musolf:2019lsf}.
The  possibility of probing the SFOEWPT-viable parameter space in the xSM  has been studied extensively (for example see~\cite{Ramsey-Musolf:2019lsf} and references therein). In the cxSM after electroweak symmetry-breaking, the model yields both a viable DM candidate (A) as well as two real neutral scalar $h_1$ and $h_2$ that are mixtures of the SM Higgs boson and the real part of the complex singlet. 
In this case, the cxSM provides more collider phenomenological signatures than xSM, such as the presence of missing transverse energy (MET) associated with pair production of A, in conjunction with decay products of one of the mixed doublet-singlet states, $h_{1,2}$. When the DM mass is 
below half that of the SM-like state $h_1$, resonant di-Higgs production may be the dominant underlying process. However, for heavier DM, there exist a variety of other subprocesses that play an important role. Thus, the SFOEWPT-viable cxSM admits a richer collider phenomenology than the xSM.

In what follows, we analyze the $b{\bar b}+\mathrm{MET}$ final state and find that it provides a powerful probe of the realization of the cxSM consistent with a SFOEWPT and DM phenomenology. 
We consider both the resonant di-Higgs portion of parameter space, wherein $m_A<m_{h_1}/2$, as well as the heavier $m_A$ regime. Present experimental constraints on $h_1$ invisible decays render the $b{\bar b}+\mathrm{MET}$ signal to be rather weak in the $m_A<m_{h_1}/2$ region~\cite{Das:2022oyx}. Consequently, we focus on the heavier $m_A$ region. We find that there exist promising prospects for cxSM discovery for DM and $h_2$ masses up to $450$ GeV and $1$ TeV, respectively.

The discussion of our analysis is organized as follows. Section~\ref{sec::model} introduces the framework of cxSM. 
Section~\ref{sec::constraints} discusses the experimental constraints on the mixing angles.
Section~\ref{sec::EWPT} describes the requirements to realise the SFOEWPT together with the cold DM candidate. 
Section~\ref{sec::DM} discusses the remaining parameter space allowed by the measurements of the DM relic density and the Higgs boson invisible decay.
Section~\ref{sec::bound_lhc} discusses the exclusion of the parameter space from the latest LHC experiments.
In section~\ref{sec::collider}, we discuss the Monte Carlo simulation of b-jets plus DM candidates in cxSM and propose a search strategy for the corresponding signals at HL-LHC.
Section~\ref{sec::conclusion} is the conclusion.
	
    	\section{The cxSM model}\label{sec::model}
The cxSM extends the SM by introducing a
complex SU(2) singlet scalar S that transforms under a global U(1) group as
$S \to Se^{i\alpha}$.
The DM candidate emerges through two ways: (a) spontaneous breaking of the global U(1) symmetry, yielding a massless Nambu-Goldstone boson; (b) inclusion of explicit, soft U(1) breaking terms in the potential, as needed to generate a DM mass. One of the two degrees of freedom in $S$ behaves like the real singlet of the xSM, and could mix with the SM Higgs boson and potentially catalyze a SFOEWPT. The other one becomes the cold DM candidate.

We consider a technically natural soft symmetry-breaking and minimal renormalizable cxSM model that do not generate additional soft symmetry-breaking terms through renormalization. 
The scalar potential at the tree-level is~\cite{Barger:2008jx}
\begin{align}\label{eq::lag}
	\nonumber
	V_0(H,S)&=\frac{\mu^2}{2} (H^\dagger H) + \frac{\lambda}{4} (H^\dagger H)^2  +\frac{\delta_2}{2} H^\dagger H |S|^2
	\\
	\nonumber
	&+ \frac{b_2}{2} |S|^2 + \frac{d_2}{4} |S|^4 \\
	&+ a_1 S + \frac{b_1}{4} S^2 +h.c..
\end{align}

The first two lines in Eq.~\eqref{eq::lag} are invariant under the U(1) transformation. 
The $a_1$ and $b_1$ terms in the third line break the U(1) symmetry explicitly. 
In general, $a_1$ or $b_1$ can be complex numbers. Under redefinition of S the quantity $\phi_S\equiv\mathrm{Arg}(b_1 a^{*2}_1$) is a rephasing invariant complex phase.
However, to obtain a viable DM candidate, mixing between the real singlet and imaginary singlet should be avoided, which requires $\phi_S=0$.
Therefore we fix $a_1$ and $b_1$ to be real numbers in the following studies.

Spontaneous symmetry-breaking (SSB) is implemented via
\begin{align}
	S&=\frac{1}{\sqrt{2}}(v_{s}+s+iA), \\
	H&=\left(
	\begin{matrix}
		G^+ \\ 
		\frac{1}{\sqrt{2}}(v_0+h+iG^0)
	\end{matrix}
	\right)\ ,
\end{align}
where $v_s$ and $v$ denote the vacuum expectation values, $G^{0,\pm}$ are the usual Higgs doublet would-be Goldstone bosons, and $s$ and $A$ denote the real and imaginary parts of the fluctuation around the singlet vacuum expectation value (vev).

Based on the U(1) symmetry breaking schemes, the model can be classified into four cases~\cite{Barger:2008jx}:
\begin{itemize}
	\item $v_s \neq 0$ and $a_1 \neq 0, ~b_1\neq0$. The  U(1) symmetry is both spontaneously and explicitly broken. We may take $\mathrm{Im}(S)$ to be the
	pseudo-Goldstone boson that is no longer massless, with its mass depending on the extent of explicit breaking via the values of
	$a_1$ and $b_1$. Note that the domain wall problem would appear if $a_1$ vanishes since a discrect $Z_2$ symmetry breaks spontaneously in this case.
	
	\item $v_s=0$ and $a_1=b_1=0$. U(1) symmetry is kept. $A$ and $s$ are identical and massive particles, such that the model is degenerate to xSM. 
	Since the U(1) symmetry is preserved, the singlet does not mix with SM Higgs and becomes two stable particles. In this case, we have two DM candidates. Comparing with the xSM, the DM relic desity is equal to twice of the xSM case.
	\item $v_s=0$ with $ ~b_1\neq0$. The U(1) symmetry is explicitly broken. The  scalar $S$ has no mixing with SM Higgs, such that $s$ and $A$ are both stable massive particles.
	Note that the $a_1$ term is  mainly to avoid a potential  domain wall problem for the case when $v_s\not=0$ as the first case. Here we can set it to be zero since we do not have SSB and, thus, no domain wall problem in this case.
	\item $v_s \neq 0$ and $a_1 = b_1= 0$. The U(1) symmetry is  spontaneously broken, yielding a massless Nambu-Goldstone boson, which we may take to be $\mathrm{Im}(S)$ and which becomes
	a possible warm DM candidate. However, such possible candidate has been ruled out suppose the warm DM candidate mass of the range $\mathcal{O}(1)$ GeV~\cite{Barger:2008jx}. 
	
\end{itemize}

In the following studies, we will focus on the most general scenario where $v_s \neq 0$ and $a_1 \neq 0, ~b_1\neq0$.

By using the minimization condition of the potential in SSB, we get

\begin{align}
	\mu^2 &= \frac{1}{2}(-v_s^2 \delta_2 - v_0^2 \lambda) \\
	\Sigma_{12} &= \frac{-4\sqrt{2}a_1-d_2 v_s^3 -v_0^2 v_s 
		\delta_2}{2 v_s},
\end{align}
where $\Sigma_{12}$ is defined as
\begin{equation}
	\Sigma_{12}=b_1 + b_2.
\end{equation}
Hence we can write down the scalar masses
\begin{equation}\label{eq::mA}
	m_A^2=-\frac{\sqrt{2}a_1}{v_s}-b_1,
\end{equation}
and
\begin{equation}\label{eq::mH}
	\mathcal{M}_h^2 \equiv 	\left(
	\begin{matrix}
		M^2_{h} & M_{hs} \\
		M_{sh} & M^2_{s} 
		
	\end{matrix}
	\right)
	=\left(
	\begin{matrix}
		\frac{1}{2}\lambda v^2_0 & \frac{\delta_2}{2} v_0 v_s \\
		\frac{\delta_2}{2} v_0 v_s & \frac{1}{2}d_2 v^2_s-\frac{\sqrt{2}a_1}{v_s}
	\end{matrix}
	\right),
\end{equation}
which can be diagonalized by an orthogonal matrix  $O(\theta)$: 
\begin{equation}
\label{eigenmassEq}
	O(\theta)^T \mathcal{M}_h^2 O(\theta) = \left(
	\begin{matrix}
		m^2_{h_1} & 0 \\
		0 & m^2_{h_2} 
	\end{matrix}
	\right),
	~~~~~O(\theta) = \left(
	\begin{matrix}
		\cos\theta & -\sin\theta \\
		\sin\theta & \cos\theta
	\end{matrix}
	\right).
\end{equation} 
Specifically, the fields are expressed in terms of mass eigenstates and the mixing angle as
	\begin{align}\label{eq::mix}
		h&=\cos\theta ~h_1 - \sin\theta ~h_2,\\ 
		s&=\sin\theta ~h_1 + \cos\theta ~h_2.
	\end{align}
The diagonal matrix, $O(\theta)^T \mathcal{M}_h^2 O(\theta)$,  gives three equations that express $\lambda$, $\delta_2$ and $d_2$ in terms of $m_{h_1}$, $m_{h_2}$, $a_1$, $v_0$, $v_s$ and $\theta$.
	\begin{align}
		\delta_2 &= \frac{\sin 2\theta ~\left(m^2_{h1}-m^2_{h2}\right)  }{v_0 v_s} \\
		\lambda &= \frac{2\left(m^2_{h_1}\cos^2\theta + m^2_{h_2}\sin^2\theta\right)}{v^2_0} \\ \label{eq::d2}
		d_2 &= \frac{2\left(\sqrt{2}a_1 + m^2_{h_2}v_s \cos^2\theta + m^2_{h_1}v_s \sin^2\theta \right)}{v^3_s}
	\end{align}
	
Meanwhile, the parameters $b_1$ and $b_2$ are related to the input parameters above and  {the DM}  mass, $m^2_A$.
	\begin{align}
		b_1 = \frac{-\sqrt{2}a_1-m^2_A v_s}{v_s}, ~~~b_2=\Sigma_{12}-b_1
	\end{align} 

So far, we have two known parameters, $v_0$ and $m_{h_1}$,  and five free parameters $m^2_A$, $m^2_{h_2}$, $a_1$, $\theta$ and $v_s$. 

Moreover, the coefficients of quartic terms should be {\it bounded from below}. We express the scalar fields as $h=\varphi \sin\alpha$ and $s=\varphi \cos \alpha$.
It is convenient to express the effective potential for a general values of $\alpha$ and $\varphi$ as~\cite{Profumo:2007wc}
	\begin{equation}
		V_{eff}(\varphi, \alpha, T)=A\varphi^4 + \bar{B}\varphi^2 + \bar{C} T^2 \varphi + D\varphi + const.,
	\end{equation}
where the $A$, $\bar{B}$, $\bar{C}$ and $D$ are massive couplings related with $v_s$, $v_0$ and T. The bar stands for that the quantity is obtained from high-T approximation. Here, the tree-level quartic coupling is
	\begin{align}
		A &= \frac{1}{16}\bigg(\lambda \cos^4\alpha + 2\delta_2 \cos^2\alpha \sin^2\alpha + d_2\sin^4\alpha \bigg).
	\end{align}
To guarantee the potential is {\it bounded from below} in any direction of $h-s$ plane, it must satisfy $\lambda > 0$, $d_2 > 0$ and $\delta_2 > -\lambda d_2$ for negative $\delta_2$. 
In addition, the requirement of positive eigenvalues of mass-squared matrix in Eq.\eqref{eq::mH} leads to $\lambda \big(d_2 - \frac{2\sqrt{2}a_1}{v^3_s}\big) > \delta_2^2$  for non-zero $a_1$~\cite{Barger:2008jx}.
	
It is useful to show the field dependent scalar masses that will be used in the calculation of the high-temperature Lagrangian.
Before electroweak symmetry breaking, the field dependent masses are
\begin{equation}\label{eq::mG}
	m_{G^{\pm,0}}^2= \frac{\partial^2 V_0}{\partial {G^{\pm,0}}^2}= \frac{1}{4} \left(2\mu^2+s^2 \delta_2 + \lambda h^2 \right),
\end{equation}
\begin{equation}\label{eq::mAf}
	m_A^2=\frac{\partial^2 V_0}{\partial {A}^2}=\frac{1}{4} \left(-4b_1 + d_2 s^2 + 2\Sigma_{12} + h^2 \delta_2 \right),
\end{equation}
\begin{equation}\label{eq::mHf}
	\mathcal{M}_h^2=\left(
	\begin{matrix}
		\frac{1}{4}\left(2\mu^2+s^2\delta_2 + 3 h^2 \lambda\right) & \frac{h s\delta_2}{2} \\
		\frac{h s\delta_2}{2} & \frac{1}{4}\left(3d_2 s^2 + 2\Sigma_{12} + h^2 \delta_2\right)
	\end{matrix}
	\right).
\end{equation}

Combining all the field dependent terms together and ignoring those field-independent terms, such as $\mu^2 T^2$ and $b_{1,2}$ terms, we could obtain the high-T approximation potential as discussed in detail in Sec.~\ref{sec::EWPT}.

    \section{Constraints on Parameters and benchmarks}\label{sec::constraints}

We first discuss the constraints of the mixing angle $\theta$ in Eq.~\eqref{eq::mix} since it is an essential parameter in cxSM for the dark matter candidate, EWPT and collider phenomenology. 
The mixing angle $\theta$ is constrained by the electroweak precision observables (EWPO) and the global Higgs measurements at LHC. 
Note that during writing this paper, the CDF experiment reported a new W mass measurement $m_W = 80.4335 \pm 0.0094$ GeV~\cite{CDF:2022hxs}, which is about 7$\sigma$ away from the SM prediction. Given that there exists some tension between this result and other experimental results, e.g. ATLAS experiment~\cite{ATLAS:2017rzl}, we prefer to not include an analysis of its implications
in this paper. We defer such an analysis to a dedicated study in the future.

\subsection{Electroweak Precision Observables (EWPO)}
The limits on the scalar mixing angle from precision electroweak measurement can be studied by assuming the extended scalar mainly contribute to the gauge boson self-energy functions.  
Modifications of the oblique parameters S, T and U~\cite{Altarelli:1997et, Peskin:1991sw} are induced due to the coupling difference between $h_1 VV$ and the SM coupling $hVV$ and due to additional contributions arising from $h_2$ via mixing.

\begin{figure}[!t]\centering
	\includegraphics[width=0.48\textwidth]{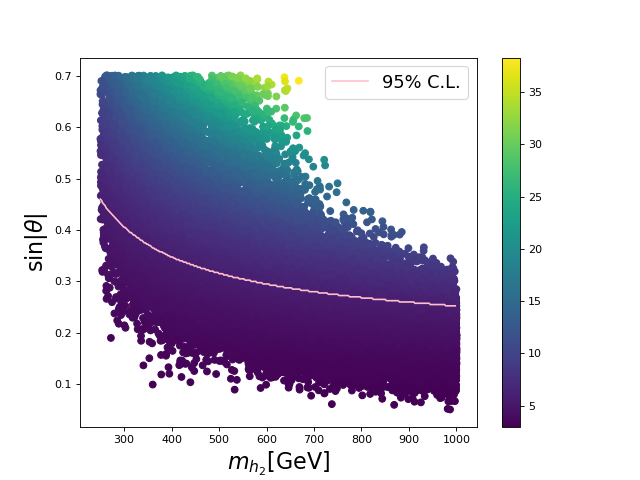}
	\caption{\label{fig::ewpo} The color bar represents the $\chi^2$ value for EWPO. The points with $\chi^2$ larger than 5.99 are excluded by EWPO. The black line corresponds to the deviation of 95\% C.L. limit. The region above the current bound is excluded.}
\end{figure}

Indeed, since the new BSM particle is a gauge singlet, no further contributions come from the gauge sector except of those associated with the mixing angle $\sin\theta$. Therefore, the deviation of EWPO operators can be expressed as~\cite{Profumo:2014opa} 
\begin{align}\label{eq::delta_stu}     \nonumber
	\Delta \mathcal{O} &= \mathcal{O}_{cxSM}-\mathcal{O}_{SM}\\ \nonumber
	&= \cos^2\theta~ \mathcal{O}(m_{h_1}) + \sin^2\theta~\mathcal{O}(m_{h_2}) - \mathcal{O}(m_{h_1})  \\
	&= \sin^2\theta \left[\mathcal{O}(m_{h_2})-\mathcal{O} (m_{h_1})\right]
\end{align}
where $m_{h_1}$ and $m_{h_2}$ are the masses of the two mass eigenstates in Eq.~\eqref{eigenmassEq} and $h_1$ is the observed Higgs boson with $m_{h_1}\approx$ 125 GeV.
Hence the deviation of a given oblique parameter $\mathcal{O}$ on EWPO from its SM value, including $\Delta S$, $\Delta T$ and $\Delta(U+S)$, is contingent upon two free parameters: $\theta$ and $m_{h_2}$. For completeness, we provide explicit expressions in terms of the Passarino Veltman functions in Appendix.~\ref{sec::appendix_EWPO}.

The best-fit values of S, T and U with respect to the SM prediction \cite{Haller:2018nnx} are
\begin{align}\label{eq::central_fit}
    \nonumber
	S-S_{SM} &= 0.04 \pm 0.11 \\ \nonumber
	T-T_{SM} &= 0.09 \pm 0.14 \\
	U-U_{SM} &= -0.02 \pm 0.11 
\end{align}

To perform the parameter scan with these experimental constraints, the $\chi^2$ is constructed as
\begin{equation}
	\chi^2 = (X-\hat{X})_i (\sigma^2)^{-1}_{ij} (X-\hat{X})_j,
\end{equation}
where the vector $X_i=(S, ~T, ~U)$ and $(X-\hat{X})_i=(\Delta S, ~\Delta T, ~\Delta U)$ are derived from Eq.~\eqref{eq::delta_stu} and defined to be the corresponding central values of the shift from SM predictions in Eq.~\eqref{eq::central_fit}.  

The quantity $\sigma^2$ is the error matrix which can be expressed as  $\sigma^2_{ij}=\sigma_i \rho_{ij} \sigma_j$. 
Here, the $\sigma_i$ is the uncertainty of $(X-\hat{X})_i$ in Eq.~\eqref{eq::central_fit}, and $\rho_{ij}$ is the correlation matrix~\cite{Haller:2018nnx} with 
\begin{equation}
	\rho_{ij}=
	\left(
	\begin{matrix}
		1 & 0.92 & -0.68 \\
		0.92 & 1 & -0.87 \\
		-0.68 & -0.87 & 1
	\end{matrix}
	\right).
\end{equation}

Fig.\ref{fig::ewpo} shows $\chi^2$ distribution of the 2-D parameter scan. The pink solid curve indicates the upper limit on the mixing angle $\sin\theta$ at 95\% C.L. as a function of $m_{h_2}$. From the plot, we could see that the mixing angle $|\sin\theta|$ is excluded  {above}  $0.35$ for $m_2 \leq$ 400 GeV and $0.25$ for $m_2 \geq$ 600 GeV. 
In the following section, we focus on the absolute value of the mixing angle lower than 0.35.

\subsection{Measurements of the Higgs boson couplings}
The mixing angle $\theta$ between $h_1$ and $h_2$ describes the coupling between SM-like Higgs boson and other SM particles and thus is constrained by measurements of the experimental measurement of the Higgs boson coupling. This section will derive the 95\% C.L. upper limit on $\sin^2\theta$ by performing a global fit to the latest LHC measurements~\cite{ATLAS:2022vkf,CMS:2022dwd}.

To characterize the impact of the cxSM on properties of the 125 GeV Higgs-like boson, it is useful to consider the signal strength, defined as
\begin{align}
	\mu_{pp\to h_1 \to XX} &= \frac{\sigma_{pp\to h_1}~BR(h_1\to X X)}{\sigma^{SM}_{pp\to h}~BR(h \to XX)_{SM}} \label{eq::mu_h1},
\end{align} 
where $\sigma_{pp\to h_1} = \cos^2\theta \times \sigma^{SM}_{pp\to h}$ is considered in tree-level. By using the decay width relationship between the SM-like Higgs and SM Higgs: $\Gamma_{h_1 \to XX} = \cos^2\theta~\Gamma_{h \to XX}$, the branching ratio of the SM-like Higgs boson decay can be expressed as: 
\begin{align}
	BR(h_1 \to XX) &=  BR(h \to X X)_{SM}. \label{eq::br_h1}
\end{align}

This equation is valid in the parameter space {relevant to the present study} , where $m_{h_2}$ is greater than $m_{h_1}$ and $m_A$ is greater than $m_{h_1}/2$. In this case, both $\Gamma_{h_1 \to A A}$ and $\Gamma_{h_1 \to h_2 h_2}$ vanish, and therefore $\mu_{pp\to h_1 \to XX}=\cos^2\theta$.

To quantify cxSM-induced deviations from SM Higgs boson properties, we construct the $\chi^2$ function for $\mu_{i\to h_1 \to f}$, where the subscript "i" stands for the production mode ({\it e.g.}, gluon-gluon fusion) and \lq\lq f\rq\rq\, indicates the decay mode:

\begin{equation}\label{eq::chi2}
	\chi^2 = \sum_{i,f} \frac{(\mu^{cxSM}_{i\to h_1 \to f}-\mu^{obs}_{i\to h_1 \to f})^2}{\sigma^2_{\mu_{i \to h \to  f}}},
\end{equation}
where all the channels tested at current LHC are considered and translated into a 95\% C.L. upper bound on $\sin^2\theta$, which translate the deviation of $\chi^2$ to be $\Delta \chi^2 \leq 3.841$. 
This is translated into an upper bound on $\sin^2\theta$, with $\sin^2\theta < 0.0468$ ($|\sin\theta|<0.216$), which is calculated based on the current global Higgs fit results:
\begin{equation}
    \mu_{\rm{ATLAS}}=1.05\pm0.06,~\mu_{\rm{CMS}}=1.02^{+0.07}_{-0.06}.
\end{equation}

    \section{SFOEWPT and Numerical Results}\label{sec::EWPT}
In this section, we consider the gauge-independent  $\mathcal{O}(T^2)$ high temperature (high-T) approximation of the finite temperature effective potential. We start with the expansion
\begin{equation}
	V_{eff}(h,s,T)=V_{0}(h,s)+V^{T=0}_{CW}(h,s)+V_\mathrm{T\neq 0}(h,s,T).
\end{equation}
$V^{T=0}_{CW}$ is the zero-temperature Coleman-Weinberg effective potential with the general form
\begin{equation}\label{eq::CW}
	V_{CW}= \sum_k \frac{(-1)^{2s_k}}{64\pi^2} g_k ~[M^2_{k}]^2\left(\log\frac{M_{k}^2}{\mu^2}+c_k\right),
\end{equation}
where $s_k$ is the spin of the $k$-th particle; $g_k$ indicates the number of degrees of freedom; $c_k$ is equal to 3/2 for scalars and fermions, and 5/6 for vector gauge bosons.

The quantity $V_\mathrm{high-T}$ is the effective potential at finite-temperature approximation at leading order in the finite temperature effective theory. It can be obtained from the conventional one-loop thermal potential
\begin{equation}\label{eq::high-t-potential}
	V_{T}^\mathrm{1-loop}=\frac{T^4}{2\pi^2}\sum_{k}n_k J_{B,F}(m_k^2/T^2)
\end{equation}

with
\begin{align}\nonumber
	J_B(\frac{m^2_k}{T^2})&=-\frac{\pi^4}{45}+\frac{\pi^2}{12}\frac{m_k^2}{T^2}-\frac{\pi}{6}\left(\frac{(m_k^2)^{3/2}}{T^3}\right)\\\nonumber
	&~~~-\frac{m_k^4}{32T^4}~\text{log}\left(\frac{m_k^2}{c_B T^2}\right)\\
	J_F(\frac{m^2_k}{T^2})&=-\frac{7\pi^4}{360}-\frac{\pi^2}{24}\frac{m_k^2}{T^2}-\frac{m_k^4}{32T^4}~\text{log}\left(\frac{m_k^2}{c_F T^2}\right), \nonumber
\end{align}
where log $ c_B=5.4076$ and log $ c_F=2.6351$. Field-dependent logarithms in $V_{high-T}$ are cancelled by $V_{CW}$ with a factor of form ln($T^2/\mu^2$) left. 
In principle, one can choose the renormalization scale to be $\mu \propto T$, so that the log-term is temperature independent.
	Moreover, at leading order in the high-temperature limit, the leading order of $V_{T\neq 0}$ is field independent and thus ignored. 
	Therefore, we keep the second order in $V_{T\neq 0}$ that is proportional to $T^2$.
	In this case, the Coleman-Weinberg potential is proportional to $M^4_k$, which is negaligable in high-T approximation, $T \gg M_k$.  
	In this paper, we use the high-T approximated potential without including the subordinate Coleman-Weinberg potential,

\begin{align}\label{eq::approx}\nonumber
	&V^{High-T}(h,s,T)\\ \nonumber
	&=V_0(h,s) +\frac{T^2}{48}\left(12m_t^2\right)\\\nonumber
	&+ \frac{T^2}{24}\left(3m_G^2+m_h^2+m_s^2+m_A^2+ 6M_W^2+3M_Z^2\right)\\ \nonumber
	&=\frac{1}{2}\left(\frac{\lambda}{8} +  \frac{\delta_2}{24} + \frac{3g^2_2 + g^2_1}{16} + \frac{y^2_t}{4}\right) h^2 T^2 \\ 
	&~~~+ \frac{\delta_2 + d_2}{48} s^2 T^2.
\end{align}
The $m_G^2$, $m_s^2$, $m_A^2$ and $m_h^2$ are field-dependent masses of the fields that interacts with the scalar fields $h$ or $s$  defined in Eq.~\eqref{eq::approx}.

Note that Eq.\eqref{eq::approx} is already gauge independent thanks to the gauge-invariant thermal masses~\cite{Patel:2011th}. Thus the critical temperature defined by high-T approximation is also gauge independent.

In the presence of the additional neutral scalar and the portal interaction, spontaneous symmetry breaking (SSB) can take place via multiple ways~\cite{Ramsey-Musolf:2019lsf}:
(a.) a single-step transition to the present pure Higgs vaccum from the symmetric phase at $T=T_{EW}$. (b.) The universe first lands in a phase with a non-zero $v_s$ at $T>T_{EW}$ followed by a transition to the current Higgs vaccum at $T_{EW}$. (c.) A one-step transition to where both the SM Higgs and the real singlet obtain vevs.
The first order EWPT can be induced at tree-level in the high-T approximated Lagrangian under certain conditions, where the situation is classified according to the number of transition steps. We discuss these possibilities below. In so doing, we first observe that a first order EWPT for scenario (a) requires that thermal loops containing the singlet scalar sufficiently enhance the term in $V_\mathrm{eff}$ proportional to $Th^3$. We do not consider this possibility here. For a discussion, see, {\it e.g.}, Ref~\cite{Ramsey-Musolf:2019lsf} and references therein.

\begin{figure}[bp]\centering 
	\begin{center}
		\includegraphics[width=0.35\textwidth]{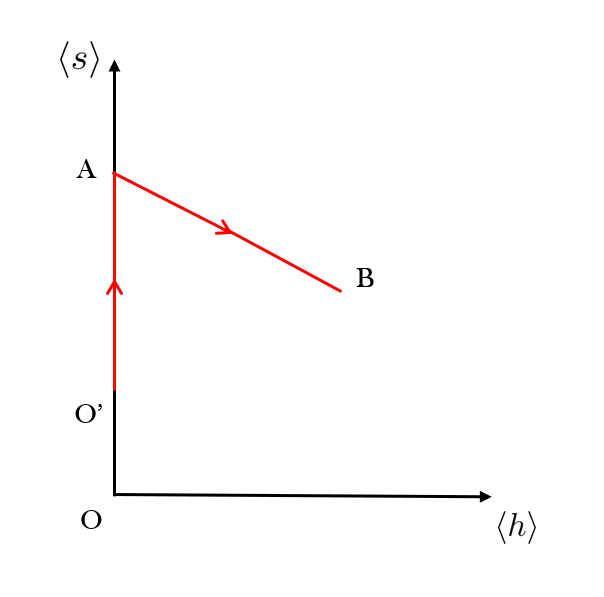}
	\end{center}
    \caption{\label{fig::EWPT} Two-step symmetry breaking at finite temperature for $a_1 \neq 0$. The first transition is continuous phase transition which occurs from $O^\prime$ to A. The second transition is from A to B, where a possible barrier can be generated between A and B for a positive $\delta_2$. Note the minimum of the potential at high temperature moves from the origin to $v_s > 0$ for a negative $a_1$.}
\end{figure}

For the two-step phase transition, as shown in Fig.~\ref{fig::EWPT}, the singlet scalar vev first moves  from $O^\prime$ to $A$, where $<s> = v_S^A/\sqrt{2} $ and $<h>=0$; the SM Higgs then also obtains its vev in the second step from $A$ to $B$, where $<s> = v_S^B/\sqrt{2} $ and $<h>=v_C/\sqrt{2} $.

For the second step, we denote the critical temperature as $T_C$, such that the strong first order electroweak phase transition can be approximated by $v_C/T_C \gtrsim 1$ with~\cite{Chiang:2017nmu}
\begin{align}
	v_C \simeq \sqrt{\frac{2\delta_2 v^A_S}{\lambda} \left(v^A_S(T_C)-v^B_S(T_C)\right)} \label{eq::vc} \\
	T_C \simeq \sqrt{\frac{1}{2\Sigma_H}\left(-\mu^2-\frac{v^A_S(T_C)^2}{2}\delta_2\right)},\label{eq::Tc}    
\end{align}
where $\Sigma_H=\frac{\lambda}{8}+\frac{\delta_2}{24}+\frac{3g^2_2+g^2_1}{16}+\frac{y^2_t}{4}$.
	
In addition, $\delta_2$ can be expressed as
\begin{equation}\label{eq::delta_2}
	\delta_2 = \frac{2}{v_0 v_s}\left(m^2_{h_1}-m^2_{h_2}\right) \sin\theta \cos\theta.
\end{equation}       
A positive $\delta_2$ can generate a barrier between two minima and therefore induce a first order EWPT, where a positive $\bar{v}^A_S-\bar{v}^B_S$ is required by Eq.~\eqref{eq::vc}. 
For the purpose of collider phenomenology, we will focus on the heavy Higgs search at HL-LHC in the following section, such that a heavy scalar resonance with $m^2_{h_2} > m^2_{h_1}$ is considered.
Thus, as implied by Eq.~(\ref{eq::delta_2})
the heavy scalar requirement requires a negative mixing angle, $ \theta$, in order for $\delta_2>0$.
Moreover, as shown in Eq.~\eqref{eq::Tc}, a positive $\delta_2$ indicates an upper limit of itself.
	
For an one-step phase transition wherein $v_0$ and $v_S$ vary from zero to nonzero at the same time, the situation is complex.
If we consider the high-T effective theory without the thermal loop-induced cubic term -- as what we performed above, such a one-step transition cannot be first order since $v_C$ is always zero. 
This can be seen from Eq.~\eqref{eq::vc} with $v^A_S$ replaced by zero. 
In principle, introducing the thermal cubic term can generate first order phase transition~\cite{Profumo:2007wc, Barger:2008jx, Chiang:2017nmu}.

With the foregoing considerations in mind, we will focus in\mrm{} this paper on the two-step phase transition.
The CosmoTransitions~\cite{Wainwright:2011kj} package is used to  numerically evaluate the EWPT quantities, e.g. $T_c$ and the corresponding vevs, and then locate the feasible parameters space for the strong first-order EWPT.
    \section{Constraints on Dark Matter Candidate} \label{sec::DM}
For the pseudoscalar $A$, since it does not mix with other scalars due to its CP-odd nature, this particle is stable and can be regarded as a dark matter candidate.
However, the $\delta_2$ term in Lagrangian generates an interaction of $g_{1AA}\cdot h_1 A A$, which can contribute to the Higgs invisible decay if the $m_A$ is less than half of the Higgs mass. Given no significant Higgs invisible decay is observed, this indicates either the coupling strength, which can be expressed as:  
\begin{equation}\label{eq::g_1AA}
	g_{h_1 AA} = \frac{\sqrt{2} a_1 + m^2_{h_1} v_s}{2 v^2_s} \sin \theta ,
\end{equation}
is highly suppressed or the $m_A$ close to or even heavier than $m_{h_1}/2$. 

To be specific, we redefine $a_1=\gamma^3 m_{h_1}^3$, ~$v_s=\beta m_{h_1}$ by introducing $\gamma$ and $\beta$, the invisible decay width can be expressed as 
\begin{align}\label{eq::higgs_dw} \nonumber
	\Gamma_{h_1 \to A A}&=\frac{g^2_{h_1 AA}}{8\pi m_{h_1}} \sqrt{1-\frac{4m^2_A}{m^2_{h_1}}}\\ 
	&=\frac{m_{h_1} }{8\pi} \left(\frac{\sqrt{2}\gamma^3+\beta}{2\beta^2}\right)^2 \sqrt{1-\frac{4m_A^2}{m_{h_1}^2}} ~\sin^2\theta \\ \nonumber
	&\sim  \left(\frac{\sqrt{2}\gamma^3+\beta}{2\beta^2}\right)^2\sqrt{1-\frac{4m_A^2}{m_{h_1}^2}}\left(\frac{\sin\theta}{0.1}\right)^2 \times 50~[\text{MeV}],
\end{align}
where the approximation in the last row is obtained by taking $|\sin \theta| = 0.1$.

The current observed upper bound on the branching ratio of Higgs invisible decay at LHC experiments is about 13\% for ATLAS~\cite{ATLAS:2022vkf} and 16\% for CMS~\cite{CMS:2022dwd}. 
While the total decay width for SM Higgs is about 4.1 MeV~\cite{CMS:2019ekd}, and notice that in Eq.~\eqref{eq::higgs_dw}, the factor $\left(\frac{\sqrt{2}\gamma^3+\beta}{2\beta^2}\right)^2 \sim \mathcal{O}(1)$ is satisfied for $\beta \sim \mathcal{O}(1)$ and $|\gamma| \sim \mathcal{O}(1)$, which indicates a narrow window for DM mass around $m_{h_1}/2$ or a delicate fine tuning cancellation between $a_1$ and $v_s$. 
Taking into account above considerations and without loss of generality, we also consider the dark matter particle $A$ in the range $60~\text{GeV} \leq m_A \leq 1$ TeV. 

Dark matter relic density and rescaled spin-independent cross section are also taken into account. To obtain the dark matter relic density, we implement the cxSM model interactions in \texttt{Feynrules}~\cite{Alloul:2013bka} to produce the \texttt{CalcHep}~\cite{Belyaev:2012qa} model file, which are then fed to \texttt{MicrOMEGAs}~\cite{Belanger:2013oya} to calculate. 
In this paper, a general scan on the free parameters is performed with
\footnote{The range of parameter values is prior and adjusted based on the scanning results. The concrete reason is given in Sec.~\ref{sec::bound_lhc}.
}
\begin{align}\label{eq::scan_space}
	\nonumber& 0\leq v_s/\text{GeV} \leq 150.0,\\
	\nonumber& |\sin\theta| \leq 0.35,\\
	\nonumber& -1000.0^3\leq a_1/\text{GeV}^3 \leq 1000.0^3,\\
	\nonumber&60.0\leq m_A/\text{GeV} \leq 1000.0,\\
	& 300.0\leq m_{h_2}/\text{GeV} \leq 1000.0.
\end{align}
In this paper, we focus on a typical process within the cxSM, the production of a pair of DM candidates associated with a SM Higgs boson, and consider it as a main concern in the definition of parameter space. We specifically select the $b\bar{b}$ decay channel of the SM Higgs, leading our search towards the $b\bar{b}$+MET channel. To enhance the $h_2\to h_1 A A$ process, we require the $h_2$ to be on-shell, with $m_{h_2}>2m_{A}+m_{h_1}\geq245\mathrm{GeV}$. Moreover, as the mass regions close to twice the SM Higgs mass are sensitive to resonant di-Higgs channels, as discussed in various studies \cite{Huang:2017jws, No:2013wsa, Chen:2014ask, Li:2019tfd, Zhang:2023jvh}, we establish a threshold to begin our scan from $m_{h_2}=300\mathrm{GeV}$. \footnote{
As for the region of $m_{h_2} \leq 60.0$ GeV, dedicated researches using the Higgs exotic decay are referred in the Refs.~\cite{Kozaczuk:2019pet, Carena:2019une, Carena:2012np, Profumo:2007wc} and references therein.
}

\begin{figure}[bp]\centering 
	\begin{center}
		\includegraphics[width=0.48\textwidth]{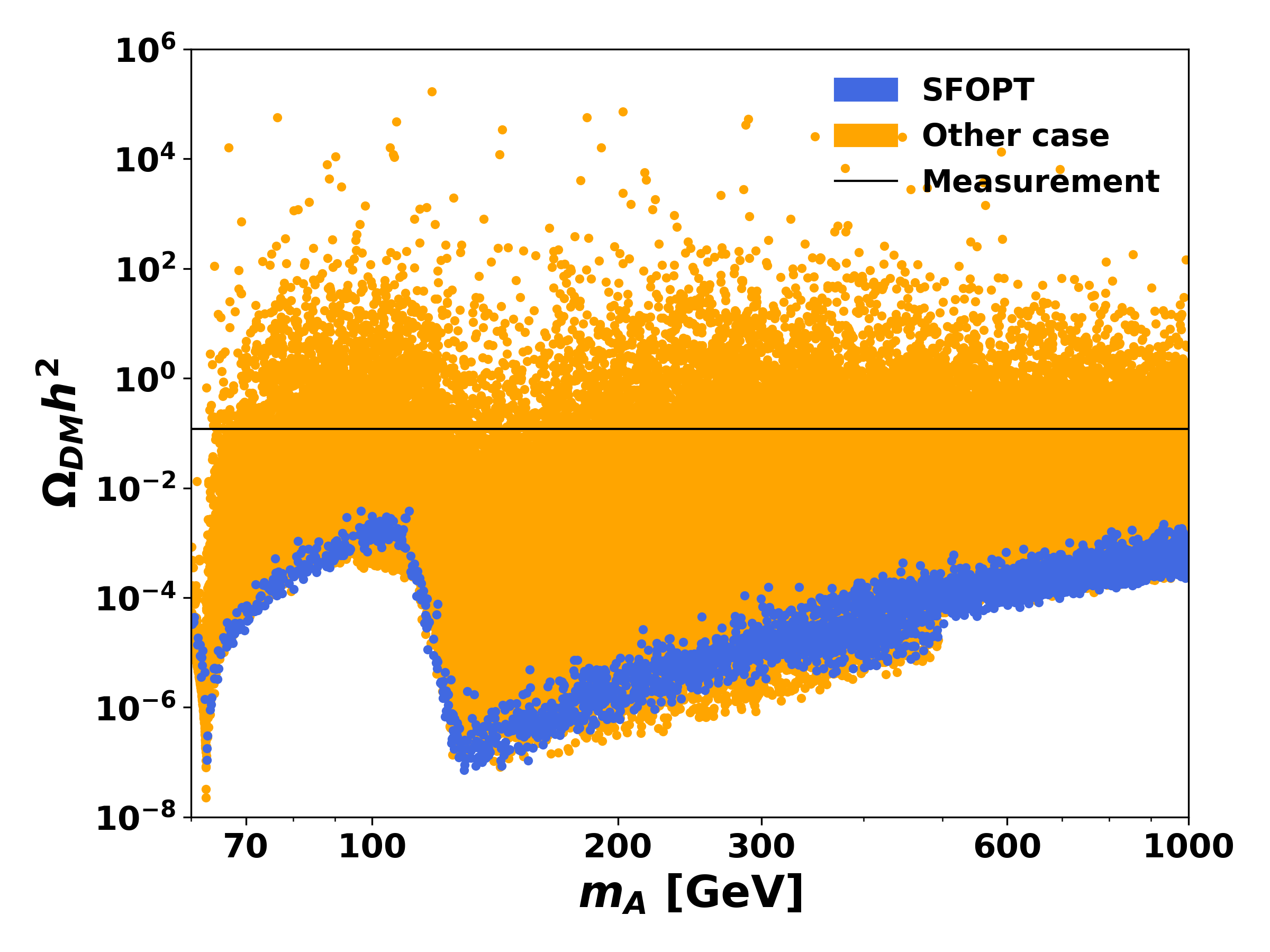}
	\end{center}
    \caption{\label{fig::DM_density} A general scan for the result of the DM relic density with the DM mass varying from 60 GeV to 1 TeV. The blue points satisfy the conditions that induce SFOEWPT, and the orange points induce second order electroweak phase transition or first order phase transition with low strength. The black solid line shows the
    cold DM relic density with $\Omega_{DM}h^2 = 0.1186$ \cite{ParticleDataGroup:2020ssz}. Valleys at 62.5 GeV and 150 GeV-500 GeV arise from DM annihilation process mediated by $h_1$ and $h_2$ respectively.}
\end{figure}

Regarding the parturbativity of the dimensionless parameters, we perform a test general scan to guarantee the dimensionless parameters satisfying a naive parturbativity constraint with $0 \leq \frac{3\lambda}{2}, \frac{\delta_2}{2}, \frac{3d_2}{2} \leq 4\pi$~\cite{Zhang:2023jvh, Gonderinger:2012rd, Chiang:2019oms}.
We assume that new physics other than the cxSM exists beyond electroweak scale, we further check such relationship at 10 TeV by solving the 1-loop RGEs. The details are presented in the appendix~\ref{sec::appdix_RGE}.

\begin{figure}[htbp]\centering 
	\begin{center}
		\includegraphics[width=0.32\textwidth]{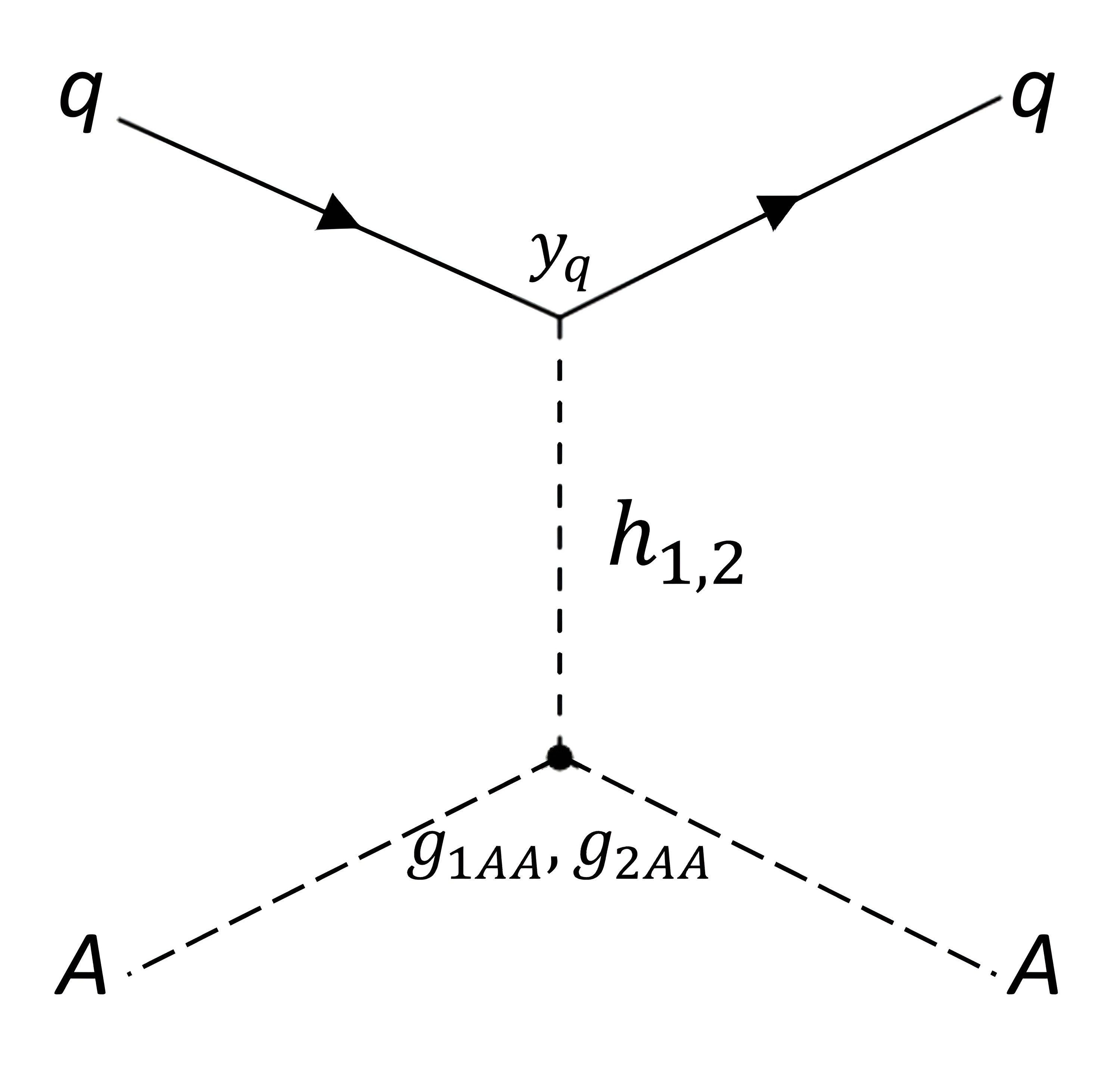}
	\end{center}
	\caption{\label{fig::DM_proton} DM direct detection: DM-proton interaction by mediating an SU(2) neutral Higgs via a t-channel process. The $h_1$ and $h_2$ interacts with the DM candidate with the nonzero $g_{1AA}$ and $g_{2AA}$.}
\end{figure}

The distribution of the DM relic density are shown in Fig.~\ref{fig::DM_density}. 
The type of EWPT and its strength in Sec.~\ref{sec::EWPT} are used to classify the points. 
The blue points represent the parameter points that could induce first order phase transition with $v_C/T_C>1$. 
The orange points contend all the other case.
Current measurement of cold DM relic density given as $\Omega_{DM}h^2 = 0.1186 \pm 0.0020$~\cite{ParticleDataGroup:2020ssz} is shown as black line. 
Most of the points in our general scan are below this line, thus satisfy the DM relic density constraint. 
There is a minimum at $m_A\simeq 62.5~\text{GeV}$ as expected where the DM annihilation process mediated by $h_1$ is highly enhanced, and valleys between $m_A\simeq 150~\text{GeV}$ and $m_A\simeq 500~\text{GeV}$ for the increase of the annihilation process mediated by $h_2$ since the scanning region of $m_{h_2}$ is chosen to be from 300 GeV to 1 TeV.

\begin{figure*}[htbp]\centering 
	\begin{center}
	\begin{subfigure}[t]{0.6\textwidth}
           \centering
		\includegraphics[width=\textwidth]{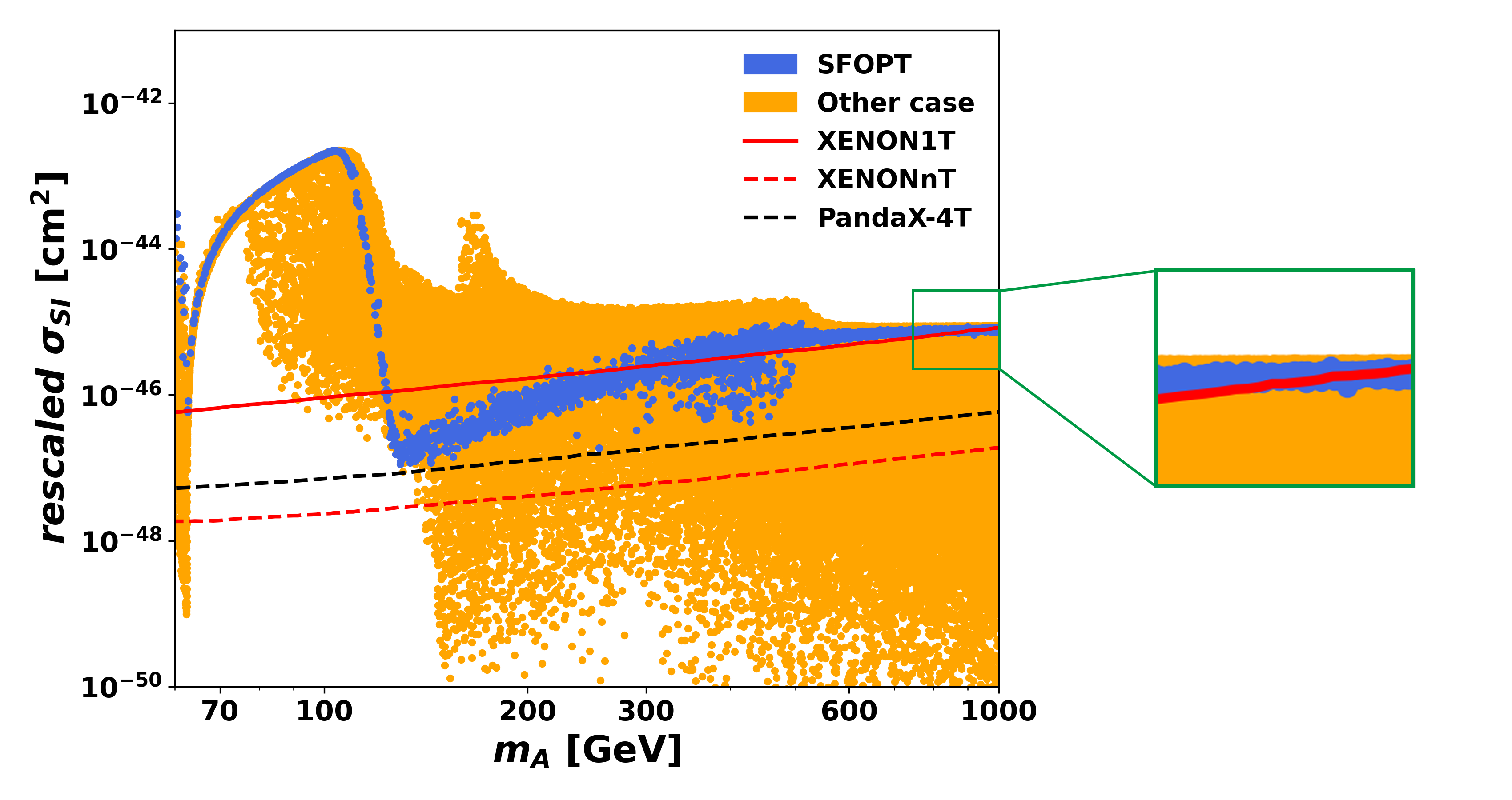}
            \caption{}
            \label{fig::DM_experiment_a}
    \end{subfigure}
	\begin{subfigure}[t]{0.39\textwidth}
           \centering
		\includegraphics[width=\textwidth]{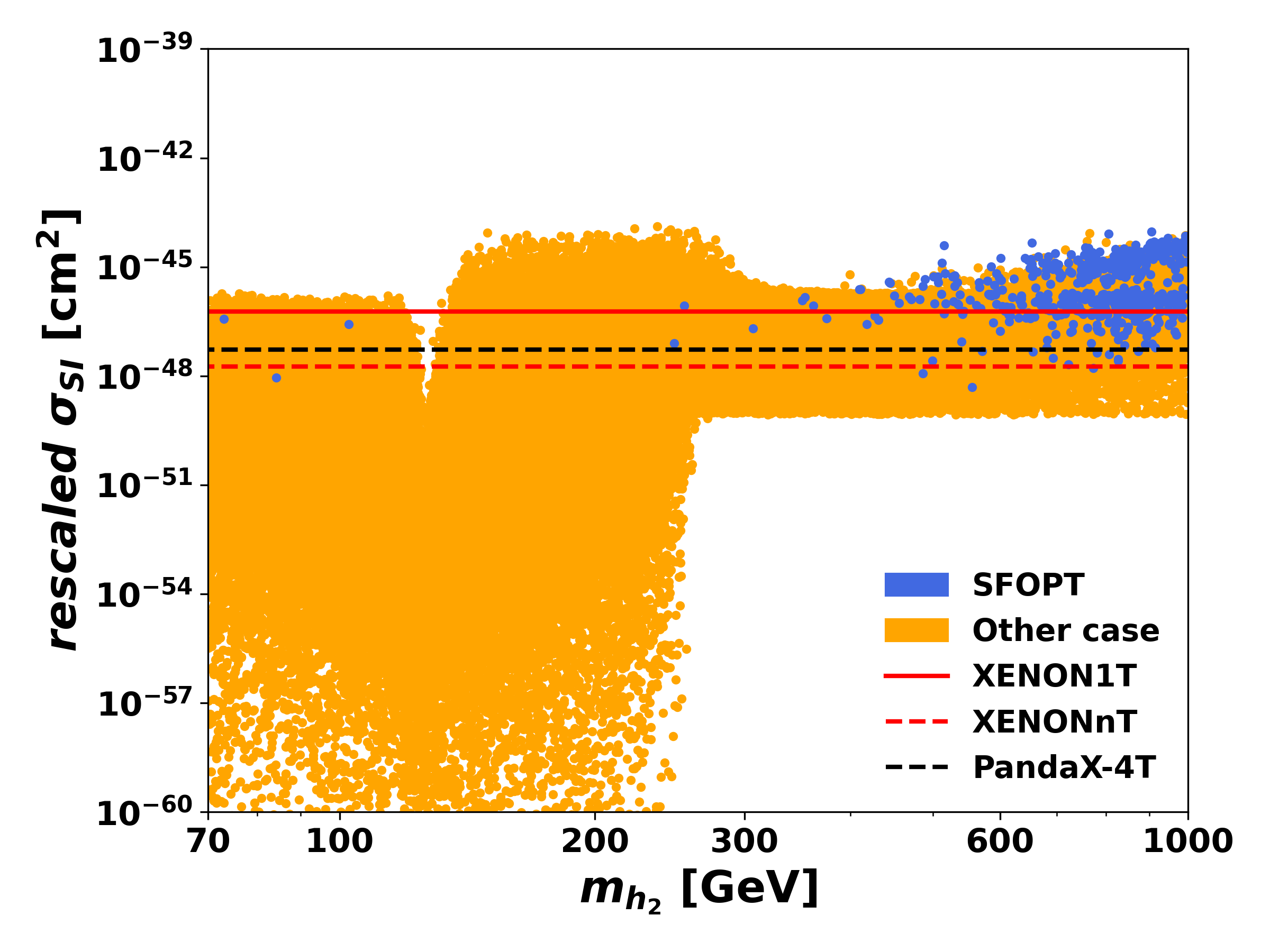}
            \caption{}
            \label{fig::DM_experiment_b}
    \end{subfigure}
	\end{center}
	\caption{\label{fig::DM_experiment}  The rescaled spin-independent DM-proton cross section of cxSM parameter points where the DM relic density is below the current measurement. Fig.(a) is the distribution of the general parameter space as shown in Eq.~\eqref{eq::scan_space}. In Fig.(b), we fix $m_A$ to 62.5 GeV and let $m_{h_2}$ vary from 70 GeV to 1 TeV. The blue points satisfy the conditions that induce SFOEWPT, the orange points induce EWPT other than strong first order. The solid line corresponds to the 95\% C.L. exclusion constrained by XENON1T and the dashed line is the expected efficiencies from XENONnT (red) and PandaX-4T (black).}
\end{figure*}

Fig.~\ref{fig::DM_proton} shows the Feynman diagram of the interaction between the dark matter particle and the proton by exchanging the SM Higgs. 	
Since the SM Higgs is composed by \eqref{eq::mix},
the spin-independent cross section of DM-proton process can be written as
\begin{align}\label{eq::sgm_si} \nonumber
	\sigma_{SI}^{[p]}&=\frac{m_p^4}{2\pi v^2(m_p+m_A)^2} \left( \frac{g_{h_1 AA}\cos\theta}{m_{h_1}^2} - \frac{g_{h_2 AA}\sin\theta}{m_{h_2}^2} \right)^2 \\
	&\times\left( f_u^{[p]} + f_d^{[p]} + f_s^{[p]} + \frac{2}{9} f_G^{[p]} \right)^2,
\end{align}
where the $f_u^{[p]} , ~f_d^{[p]}, ~f_s^{[p]} $ and $ f_G^{[p]}$ are proton form factors \cite{Belanger:2013oya} and the minus sign in the first bracket is derived from the minus sign in Eq.~\eqref{eq::mix} with the couplings being
\begin{align}
	g_{h_2 AA}&=\frac{\sqrt{2}a_1+m_{h_2}^2 v_s}{2 v_s^2} \cos\theta, \label{eq:g2AA} \\ 
	g_{h_1 AA}&=\frac{\sqrt{2}a_1+m_{h_1}^2 v_s}{2 v_s^2} \sin\theta. \label{eq::g1AA}
\end{align}
	
In this work, \texttt{MicrOMEGAs}~\cite{Belanger:2013oya} is also used to calculate the spin-independent cross section. If the DM abundance is less than the observed DM abundance, the rescaled spin-independent cross section $\sigma_{SI}\text{(rescaled)}$ could be obtained according to
\begin{equation}
	\sigma_{SI} \text{(rescaled)} = \sigma_{SI} \frac{\Omega_{cxSM}h^2}{\Omega_{DM}h^2}.
\end{equation}

The general scan with Eq.~\eqref{eq::scan_space} is also performed to $\sigma_{SI}\text{(rescaled)}$ as shown in Fig.~\ref{fig::DM_experiment_a}.
The definition of color remains the same as that in the figure of dark matter relic density. 
Experimental constraints from the direct dark matter search experiment XENON1T~\cite{XENON:2018voc} is shown as line, and the expected efficiencies of future experiment XENONnT~\cite{XENON:2020kmp} and PandaX-4T~\cite{PandaX-4T:2021bab} are shown by the dashed line.
\footnote{Results that are close to the XENON1T constraint can be obtained from current new experiments such as the LUX-ZEPIN~\cite{LZ:2022ufs}.}
Currently we can exclude the dark matter mass between 65 GeV and 120 GeV under the premise of SFOEWPT. Most of the SFOEWPT points in our scanning space can be covered by XENONnT.

Fig.~\ref{fig::DM_experiment_b} shows the scaled cross section v.s. singlet-like Higgs mass with the $m_A$ fixed to 62.5 GeV and $m_{h_2}$ Varying from 70 GeV to 1 TeV. A minimum is generated when $m_{h_2}=m_{h_1}$ as indicated in Eq.~\eqref{eq::sgm_si}. From Fig.~\ref{fig::DM_experiment_b}, we can see that very few parameter points for SFOEWPT can survive from the direct dark matter search. 
On the contrary, in the Fig.~\ref{fig::DM_experiment_a} most parameter regions with $m_{A}>62.5 ~\mathrm{GeV}$ that realise SFOEWPT survive the current direct DM search and are able to be tested by XENONnT.
Therefore, it is more valuable for DM direct detection to investigate the $m_A$ region beyond 62.5 GeV.

A similar study on the DM relic density is presented in the Ref.~\cite{Chen:2019ebq}, which, same as this paper, suggests that most of the parameter region satisfying DM relic density, and SFOEWPT conditions survives the Xenon-1T search and can be probed by Xenon-nT and PandaX-4T.
Compared with Ref.~\cite{Chen:2019ebq}, this paper finds some parameter space that survives Xenon-nT search. We further study the cases of $m_A\simeq62.5$ GeV and $m_A>62.5$ GeV, and reaches the SFOEWPT parameter region beyond the detection capability of XENONnT.

    \section{Heavy scalar resonance searches bounds at the LHC}\label{sec::bound_lhc}

\begin{figure*}[htbp]\centering 
    \begin{center}
	\begin{subfigure}[t]{0.305\textwidth}
           \centering
		\includegraphics[width=1\textwidth]{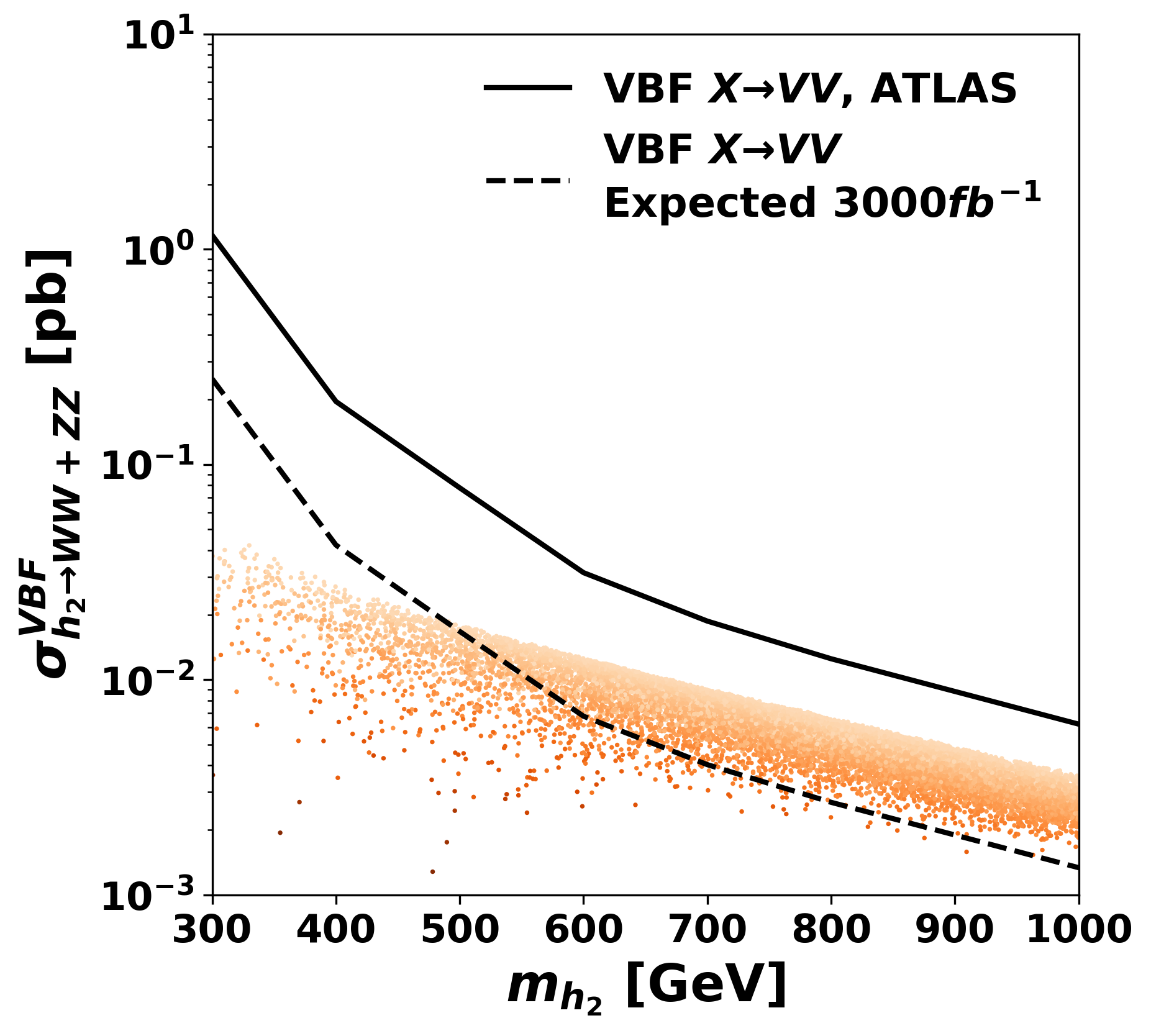}
            \caption{}
            \label{fig::scalar_ATLAS_a}
    \end{subfigure}
	\begin{subfigure}[t]{0.305\textwidth}
           \centering
		\includegraphics[width=1\textwidth]{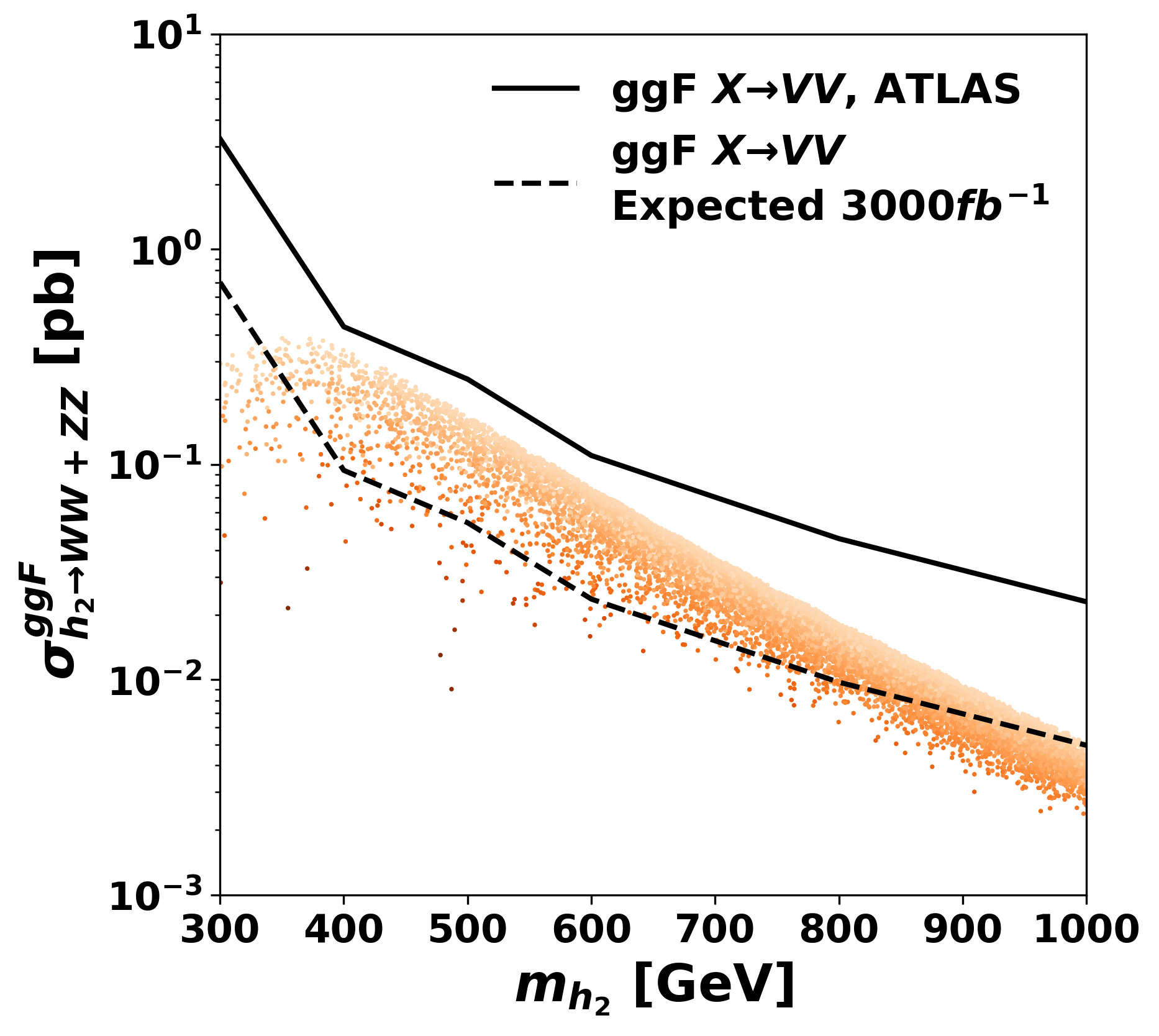}
            \caption{}
            \label{fig::scalar_ATLAS_b}
        \end{subfigure}
	\begin{subfigure}[t]{0.37\textwidth}
           \centering
		\includegraphics[width=1\textwidth]{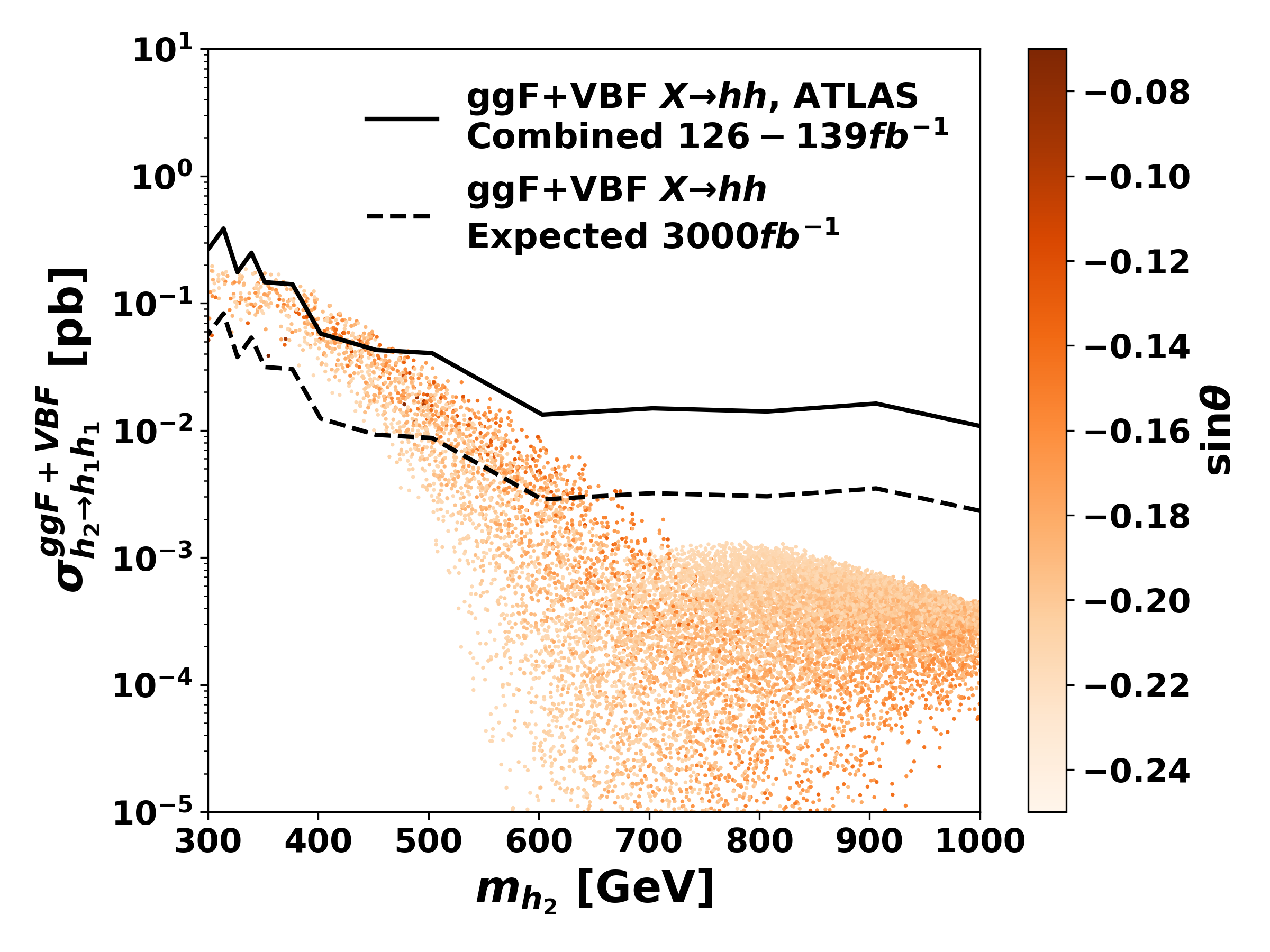}
            \caption{}
            \label{fig::scalar_ATLAS_d}
    \end{subfigure}
	\end{center}
	\caption{\label{fig::scalar_ATLAS} Cross sections of SFOEWPT parameter points for (a) $\mathrm{VBF}~h_2 \to VV$, (b) $\mathrm{ggF}~h_2 \to VV$ and (c) $\mathrm{VBF+ggF}~h_2 \to h_1 h_1$ as functions of $m_{h_2}$.  
	The colorbar represents the mass of pseudoscale boson $A$. 
	The black curves show the 95\% C.L. upper limit from ATLAS heavy resonance searches \cite{ATLAS:2020fry,ATLAS:2021tyg}, above which the parameter points are excluded. }
\end{figure*}

The cxSM predicts that the singlet-like scalar boson $h_2$ can be produced at the LHC and decay to various standard model particles. Thus, $h_2$ can behave as a heavy spin-0 resonance in collider when $m_{h_2}>m_{h_1}$.
In this section, we investigate the constraints on the cxSM parameter space from the direct heavy resonance search at the LHC.
The production cross section times branching fraction of $h_2 \to WW$~\cite{ATLAS:2020fry}, $ZZ$~\cite{ATLAS:2020fry}, $hh$~\cite{ATLAS:2021tyg}, $\tau\tau$~\cite{ATLAS:2020zms}, $bb$~\cite{ATLAS:2019tpq} and $tt$~\cite{ATLAS:2022rws} are scanned in the parameter space Eq.~(\ref{eq::scan_space}).

These calculations rely on the mixing angle $\theta$ and the widths of additional decay. Given by Eq.~\eqref{eq::mix}, the production cross section of $h_2$ can be expressed as $\sigma_{p p \to h_2} = \sin^2{\theta} ~\sigma_{p p \to h}$ for each production mode. The decay widths of the existing channels are also obtained by multiplying a factor $\sin^2{\theta}$ on the standard model widths as $\Gamma_{h_2\to XY}=\sin^2{\theta}\Gamma_{h\to XY}^{SM}$. The Standard Model cross sections and decay widths are obtained from CERN Yellow Report 4~\cite{LHCHiggsCrossSectionWorkingGroup:2016ypw}. For the rare decay channels, the $h_2\to AA$ decay is considered because of the $\delta_2$ term in Lagrangian, similar to the discussion of $h_1 \to AA$ in Sec.~\ref{sec::DM}. The $h_2 h_1 h_1$ vertex also exists with the coupling 
\begin{align} \nonumber
    g_{h_2 h_1 h_1} = &\sin{\theta}\cos{\theta}\times\\
    &\left[\frac{3a_1}{\sqrt{2}v_s}\frac{\sin{\theta}}{v_s}+(m_{h_1}^2+\frac{m_{h_2}^2}{2})(\frac{\sin{\theta}}{v_s}-\frac{\cos{\theta}}{v_0})\right],
\end{align}
due to the Higgs sector mixing. Thus, we must also include the $h_2\to h_1 h_1$ channel. In addition, the three-body decay channel, $h_2 \to h_1 A A$, is also taken into consideration because of the non-zero coupling of $g_{h_2 h_1 AA}$ with
\begin{equation}\label{eq::coupling_4}
\begin{split}
         g_{h_2 h_1 AA}&=\frac{1}{2v_0v_s^3}(\sqrt{2}a_1 v_0 \sin\theta \cos\theta+ m^2_{h_2} v_s^2 \cos^2\theta \sin^2\theta \\
         &- m_{h_1}^2 v_s^2 \cos^2\theta \sin^2\theta + m_{h_1}^2 v_s v_0 \cos\theta \sin^3\theta\\
         &+ m_{h_2}^2 v_0 v_s \cos^3\theta \sin\theta).
\end{split}
\end{equation}
Apart from the direct $h_2\to h_1 A A$ decay, an interesting process where one or both of the Higgs boson from di-Higgs decay channel is off shell, leading to one or more pairs of heavy particles ($WW$, $t{\bar t}$ {\it etc} or a pair of heavy dark matter particles) in the final state, e.g. $h_2\to h_1 h_i^\ast\to h_1 A A$. One nominally expects these contributions to be suppressed due to the off-shell $h_1$ propagator and additional-particle phase space suppression. We find, however, that the contribution from the $h_2\to h_1 h_i^\ast\to h_1 AA $ channel can provide significant discovery potential. The differential cross section of the mediate three-body decay process is calculated according to the Appendix.~\ref{sec::appdix_3_body}, by integrating which we can obtain the width.



With these additional decays, the branching ratio for a decay from $h_2$ to standard model particles can be written as

\begin{equation}
	BR(h_2 \to XX)=\frac{\sin^2{\theta}~\Gamma_{h \to XX}}{\sin^2{\theta} ~\Gamma^{SM}_{h} + \Gamma^{BSM}_{h_2}},
\end{equation}
where 
\begin{equation}\label{eq::gamma_bsm}
    \Gamma^{BSM}_{h_2}=\Gamma_{h_2 \to h_1 h_1}+\Gamma_{h_2 \to AA} + \Gamma_{h_2 \to h_1 AA}+\Gamma_{h_2 \to h_1 t\bar{t}}.
\end{equation}

\begin{figure*}[htbp]
    \centering 
    \begin{center}
        \begin{subfigure}[t]{0.24\textwidth}
            \centering
            \includegraphics[height=4.1cm]{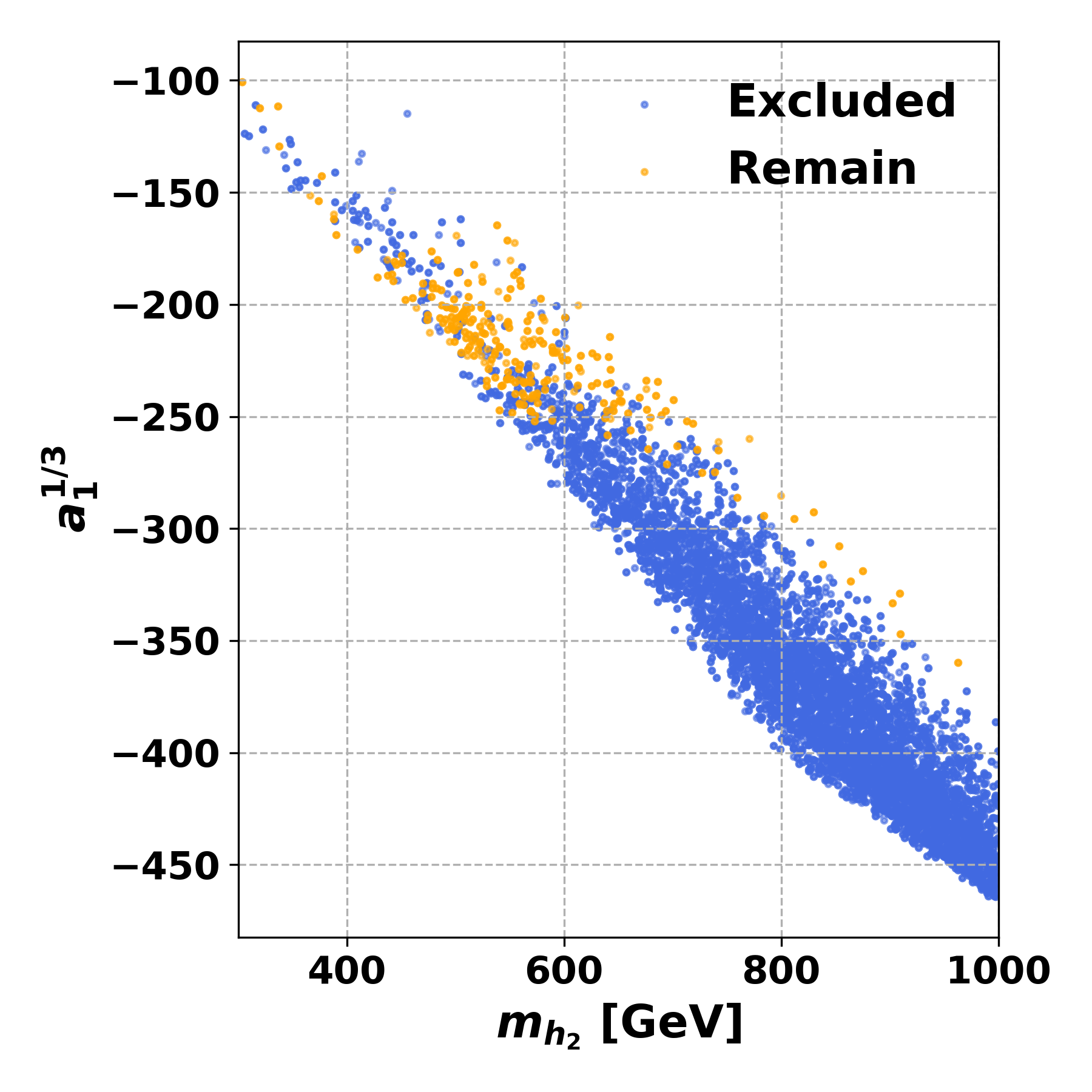}
            \caption{}
            \label{fig::correlation_a1}
        \end{subfigure}
        \begin{subfigure}[t]{0.24\textwidth}
            \centering
            \includegraphics[height=4.1cm]{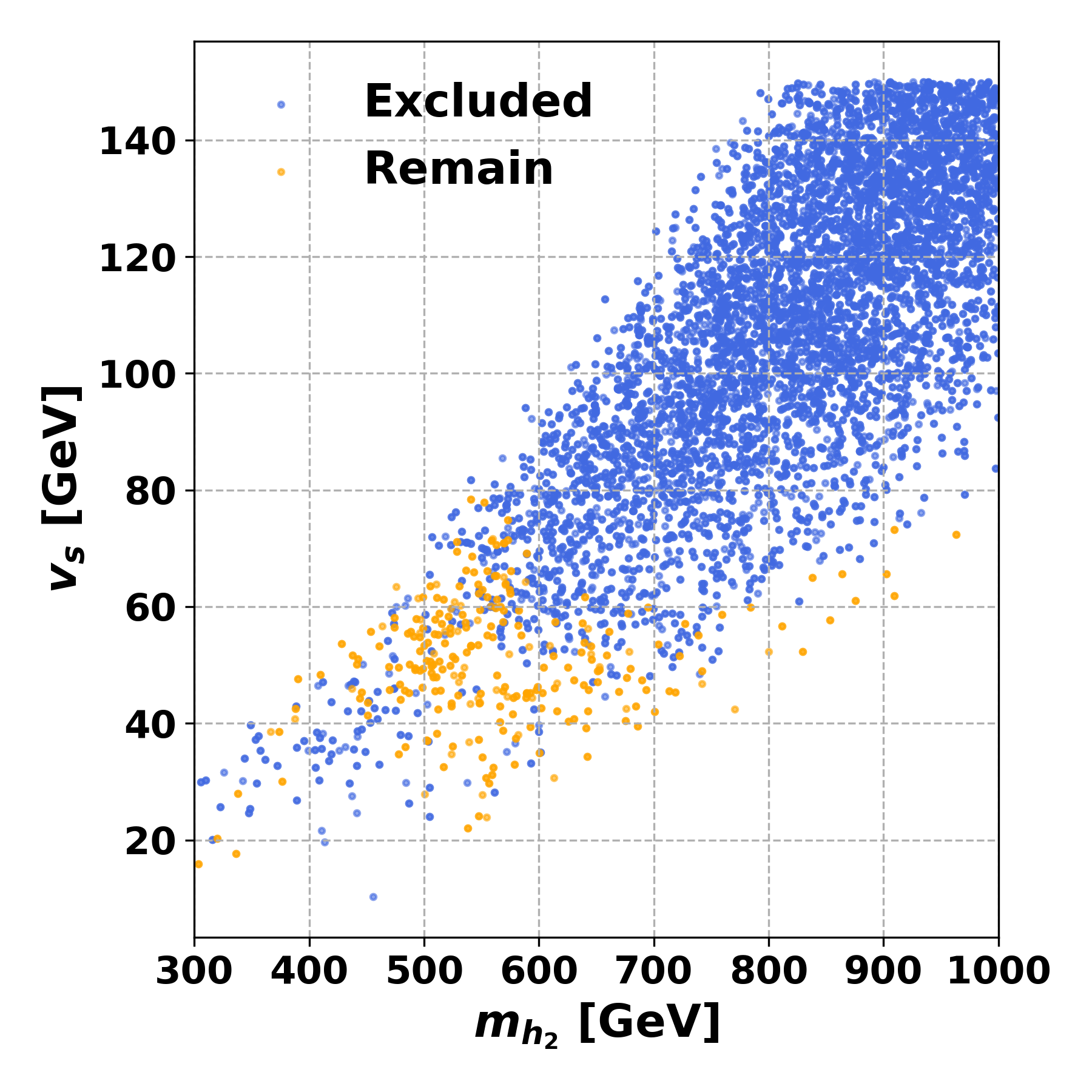}
            \caption{}
            \label{fig::correlation_vs}
        \end{subfigure}
        \begin{subfigure}[t]{0.24\textwidth}
            \centering
            \includegraphics[height=4.1cm]{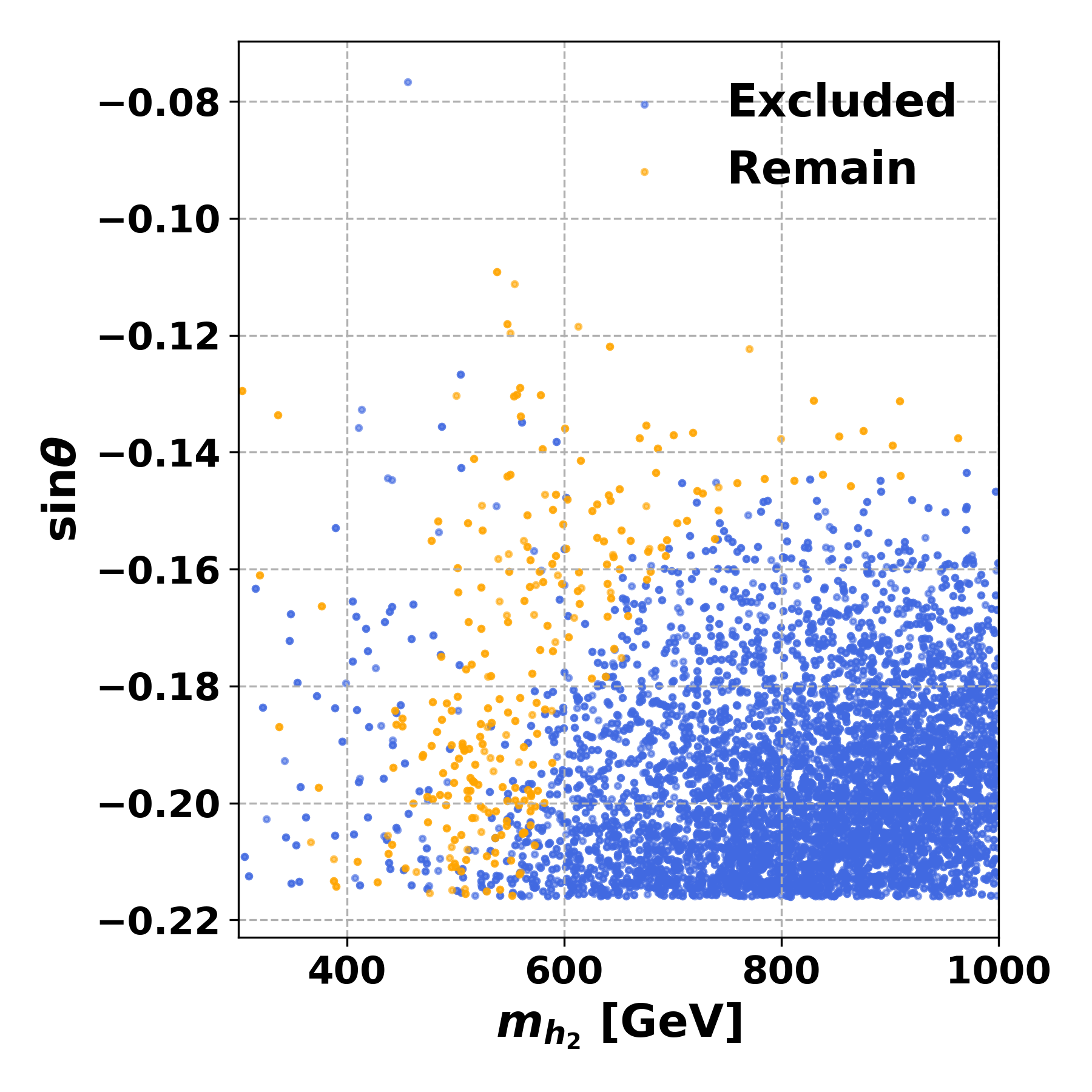}
            \caption{}
            \label{fig::correlation_sin}
        \end{subfigure}
        \begin{subfigure}[t]{0.24\textwidth}
            \centering
            \includegraphics[height=4.1cm]{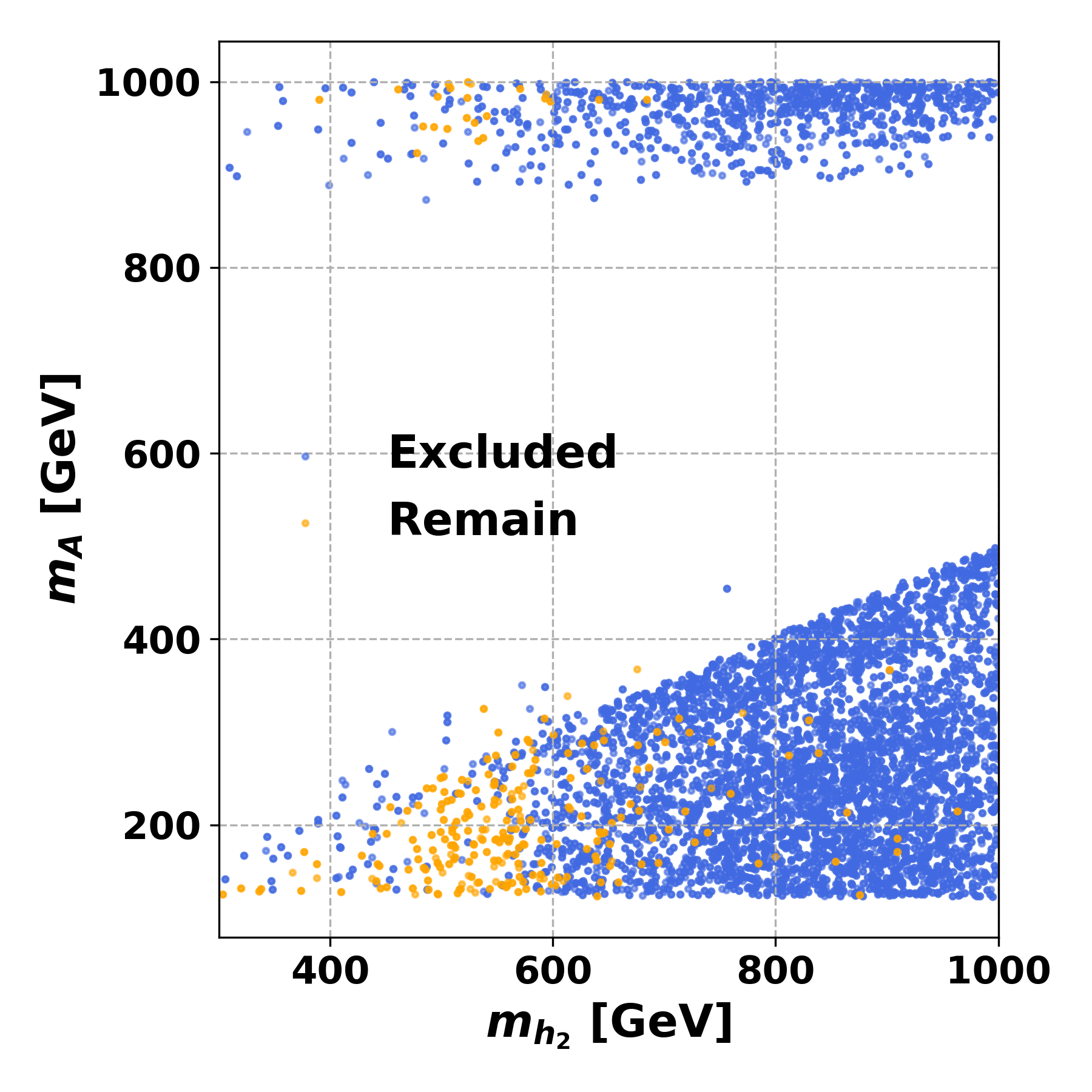}
            \caption{}
            \label{fig::correlation_mA}
        \end{subfigure}
    \end{center}
    \caption{\label{fig::correlations} Relationships between $m_{h_2}$ and $a_1$, $m_{h_2}$ and $v_s$, $m_{h_2}$ and $\sin\theta$, and $m_{h_2}$ and $m_A$ respectively, after applying all the current constraints. The blue points represent parameter points eligible for investigation through direct resonance searches at the HL-LHC, and the yellow points are expected to persist even after this search.}
\end{figure*}

Finally, the overall cross section in cxSM for heavy resonance search can be simply written as $\sigma_{pp\to h_2}\times BR(h_2 \to XX)$. 
Fig.~\ref{fig::scalar_ATLAS_a} and \ref{fig::scalar_ATLAS_b} depict experimental constraints from $h_2 \to WW$ and $ZZ$ decay channels for parameter points satisfying SFOEWPT. Both vector-boson fusion (VBF) and gluon-gluon fusion (ggF) production modes are considered. The black curves in the figures are the experimental upper limit on the overall cross section, above which the parameter points are excluded. 
For the heavy resonance search at the HL-LHC, the expected efficiency can be estimated by the current upper limit and a factor of $1.18 \times \sqrt{3000.0/139.0}$. The factor of 1.18 is the ratio of the 14 TeV LHC and 13 TeV LHC cross sections~\cite{Zhang:2023jvh}. The second factor comes from the integrated luminosity. The corresponding upper limit is given by the dashed line in the Fig.~\ref{fig::scalar_ATLAS_a} and Fig.~\ref{fig::scalar_ATLAS_b}.
One can find that even though most of the viable SFOEWPT parameter space survives from the LHC Run-2 searches, the forthcoming Run-3 with 3000 fb$^{-1}$ integrated luminosity manifests powerful capacity in detecting the singlet induced SFOEWPT, which is consistent with the prediction in the Ref.~\cite{Zhang:2023jvh}. Note that the di-boson channel is most powerful in heavy resonance mass region where $m_{h_2} > 500$ GeV. For $m_{h_2} \lesssim 500$ GeV, a considerable portion of the SFOEWPT- as well as DM-viable space cannot be probed.

Fig.~\ref{fig::scalar_ATLAS_d} presents constraints for the same points from ggF+VBF di-Higgs searches, combining results from $b\bar{b}b\bar{b}$, $b\bar{b}\gamma\gamma$, and $b\bar{b}\tau\bar{\tau}$ final states. 
Unlike the di-boson case, the di-Higgs channel manifests relatively powerful detection ability in probing $m_{h_2} \lesssim 500$ GeV signals, whereas the capability around $\sim 500$ GeV is limited.
Other channels, including $h_2 \to\tau\tau$, $h_2 \to t\bar{t}$ and $h_2 \to bb$, are found to negligible exclusion power in the scanned parameter space and thus not shown in the figures. 
Those points with heavy $h_2$ that survive the di-boson searches, $h_2\to VV$, are likely to have lower $\sin\theta$ and lower $A$ mass. This is because the BSM branching ratio $h_2\to AA$ ($h_2\to h_1 AA$) becomes nonzero for $m_{h_2}\geq 2 m_A$ ($m_{h_2}\geq 2 m_A+m_{h_1}$) and thus reducing the branching ratio for $h_2 \to VV$ and making it more difficult for experiment to exclude this space via di-boson resonance searches.
Importantly, without consideration of the $b{\bar b}$+MET signature, none of the foregoing channels can distinguish whether the signal comes from the xSM or a cxSM.

The correlations between $m_{h_2}$ and $v_s$, $\sin\theta$, and $m_A$ are used to explain the range of parameter values chosen in Sec.~\ref{sec::DM} and are depicted in Fig.~\ref{fig::correlations}. 
All points satisfy SFOEWPT, DM and LHC Run-2 constraints. 
The blue ones are able to be probed by the future HL-LHC at the $2\sigma$ level, while the yellow ones will survive the HL-LHC searches. The plots represent the parameter distribution in the SFOEWPT-viable cxSM and reflect our strategies for selecting the parameter scanning range:

\begin{itemize}
    \item The mixing angle that is proportional to $v_s$ and expressed by
          \begin{equation}
             \sin 2\theta = \frac{\delta_2 v_0 v_s}{m_{h_1}^2 - m_{h_2}^2},
          \end{equation}
          is highly constrained by EWPO, the Higgs measurement as well as the di-boson searches, which, therefore, sets an upper limit to the vev of the singlet. Notice that in the region $m_{h_2}\lesssim 500$ GeV, no obvious correlation with $\sin\theta$ is found, as indicated in Fig~\ref{fig::correlation_sin}. This absence is due to the weak detection capability in this region  from the di-boson channels.
          On the other hand, a value of $\sin\theta$ approaching zero is incompatible with SFOEWPT. This is due to the necessity of having a non-negligible value of $|\delta_2|$, as indicated in Eq.~\ref{eq::delta_2}, to meet the criterion of $v_C/T_C\gtrsim 1$.
    
    \item The parameter range of $m_{h_2} \leq 1$ TeV is chosen because rare SFOEWPT-viable parameter space that satisfies HL-LHC constraints and DM constraints is seldom found beyond 1 TeV. Note that this rough upper bound is consistent with the general arguments in Ref.~\cite{Ramsey-Musolf:2019lsf}.

    \item In Fig.~\ref{fig::correlation_a1}, the linear-like relationship between $a_1^{1/3}$ and $m_{h_2}$ is due to the perturbation requirements on $0\leq d_2\leq 8/3\pi$, where $d_2$ in Eq.~\ref{eq::d2} can be expressed as:
        \begin{equation}
             d_2 = \frac{2}{v_s^3}\left[m_{h_1}^2 v_s+
             (m_{h_2}^2-m_{h_1}^2)v_s \cos^2\theta +\sqrt{2}a_1\right].
        \end{equation}
    As $m_{h_2}$ increase, a more negative value of $a_1$ is needed to offset the contribution from $m^2_{h_2}$.  Furthermore, $a_1$ is pivotal in shifting the vacuum expectation value in the $<s>$ direction before the second step EWPT. A larger magnitude of $|a_1|$ is generally advantageous for facilitating strong first-order phase transitions~\cite{Chen:2019ebq}.
    
    \item The $m_A$ distribution falls into two separate areas. This is because the range of values for  $m_A$ that do not exceed the XENON-1T DM direct detection upper limit is discontinuous, as indicated in Fig.~\ref{fig::DM_experiment}. Notice that under the XENON-1T exclusion line, there exists few SFOEWPT-viable parameter space around $m_A \sim$1 TeV, which is the reason for the discontinuity in $m_A$. 
\end{itemize}

\begin{figure*}[t]\centering 
		\begin{center}
           \centering
		\includegraphics[width=1\textwidth]{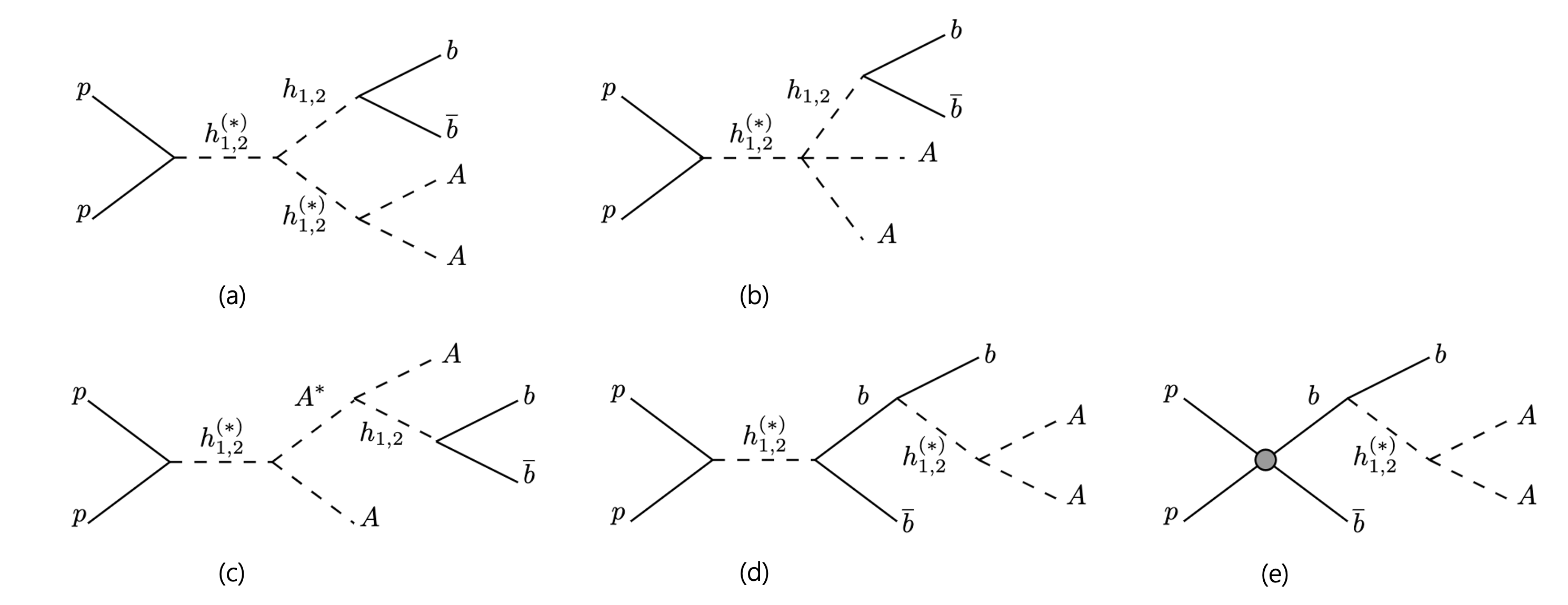}
		\end{center}
  \caption{\label{fig::feyn_diag}Representative Feynman diagram to generate signal events with b-jets plus MET final states at the LHC. }
\end{figure*}

\section{Prospect of heavy scalar search in b-jets+MET channels}\label{sec::collider}

When producing the cxSM $b{\bar b}+\mathrm{MET}$ signal, we consider a comprehensive set of processes (CSPs) that contribute to this channel.
Our search strategy is inspired by strategies used for mono-Higgs plus MET, then optimized to account for other important sub-processes, such as those in which an off-shell $h_1$ mediates $b{\bar b}$ pair production.
To carry out detailed simulations for the HL-LHC, we select a set of benchmark parameter points after applying all the constraints and requirements discussed in the previous sections. In subsection~\ref{subsection::CSPs}, we explore aspects of the underlying sub-processes and allowed parameter space, as it bears on the LHC signal.
The selection criteria and the signal signature are shown in subsection~\ref{subsection::analysis}. 
Finally, we find that the discovery potential with a significance of $\geq 1.96\sigma$ reach for the $b\bar{b}+\text{MET}$ channel is significant at the HL-LHC, and most parameter points will be covered in that case.

\subsection{The complete set of cxSM processes for b-jets plus MET}\label{subsection::CSPs}

In the cxSM, multiple processes contribute to the $b\bar{b}+\text{MET}$ final state, including the di-Higgs channels, heavy Higgs boson direct decay channels and mono-Higgs plus b-jets.
The DM candidate can be produced from direct four-particle vertex from heavy Higgs boson $h_2$ or from the subsequent decay of an on-shell or off-shell $h_{1,2}$ boson.
We consider all the processes with the coupling order satisfying $\text{QCD}\leq 2$ and $\text{QED} \leq 4$ in MadGraph~\cite{Alwall:2011uj}.
The CSPs have more than one hundred diagrams. A brief overview of the main types is illustrated in Fig.~\ref{fig::feyn_diag}, among which the cross section is dominated by the diagram-\ref{fig::feyn_diag}(a) and diagram-\ref{fig::feyn_diag}(b), in particular, the diagram-\ref{fig::feyn_diag}(b) with mediator substituted by an off-shell $h_{1}$ is found to be significant. 

Previous studies on the collider searches of the cxSM include:
\begin{itemize}
    \item The $h_1\to A~A$ case with $m_{A}=62.5$ GeV~\cite{Chiang:2017nmu}, which satisfies the Higgs invisible decay constraint and obtains a relatively large parameter space.
    \item The degenerate-scalar scenario with $|m_{h_2}-m_{h_1}|\lesssim \mathcal{O}(1)$ GeV~\cite{Abe:2021nih, Cho:2022zfg}.  
    Collider signatures in this scenario are SM-like, and therefore current experimental data cannot distinguish them from the SM predictions.
\end{itemize}

However, the on-shell $h_1\to A~A$ decay with $m_A= 62.5$ GeV is not expected to significantly enhance the sensitivity of $b\bar{b}+$MET search because the branching ratio is already highly bounded by the Higgs invisible decay constraint. Moreover, with $m_A= 62.5$ GeV, we find that the parameter space is tightly constrained by the current experimental requirements. Therefore, in this study, we investigate the most general case where $m_A \geq 62.5$ GeV. However, due to the exclusion of all points with $m_A$ in the range of [62.5 GeV, 120 GeV] by XENON1T, as mentioned in Sec.~\ref{sec::DM}, we further restrict our analysis to $m_A \geq 120$ GeV. 

To choose benchmark mass points for analysis, we impose a requirement that $m_{h_2} > m_{h_1} + 2 \times m_A$. This condition ensures that the $h_2$ mediator in diagram-\ref{fig::feyn_diag}(a) and diagram-\ref{fig::feyn_diag}(b) can be on-shell, and thus enhances the cross section of CSPs signal. Therefore, the analysis will be conducted on the following ten mass points:

\begin{table}[htbp]
  \centering
  \resizebox{\textwidth/2}{!}{
    \begin{tabular}{|c || c c c c c c c c c c|}\hline
     $m_A$/GeV & 130 & 130 & 130 & 130 & 230 & 230 & 230 & 330 & 330 & 430\\
     \hline
     $m_{h_2}$/GeV & 400 & 600 & 800 & 1000 & 600 & 800 & 1000 & 800 & 1000 & 1000\\ \hline
     \end{tabular}
  }
  \caption{Mass points used to analyze.}
  \label{tab:mass_points}
\end{table}

Taking into account all the current constraints and requirements discussed in the previous sections, it is impossible to find a shared benchmark point for the remaining parameters ($a_1$, $v_s$, $\sin\theta$) that satisfies all the mass points. For instance, The SFOEWPT tends to favor a larger $-a_1$ as $m_{h_2}$ becomes heavier. The relationship between $m_{h_2}$ and $a_1$ is depicted in Fig.~\ref{fig::correlation_a1}, from which it is evident that there is no single choice for $a_1$ that can be used for the mass range  between $m_{h_2}=400$ GeV and $m_{h_2}=1000$ GeV.

\begin{figure}[!htbp]\centering 

        \includegraphics[width=0.45\textwidth]{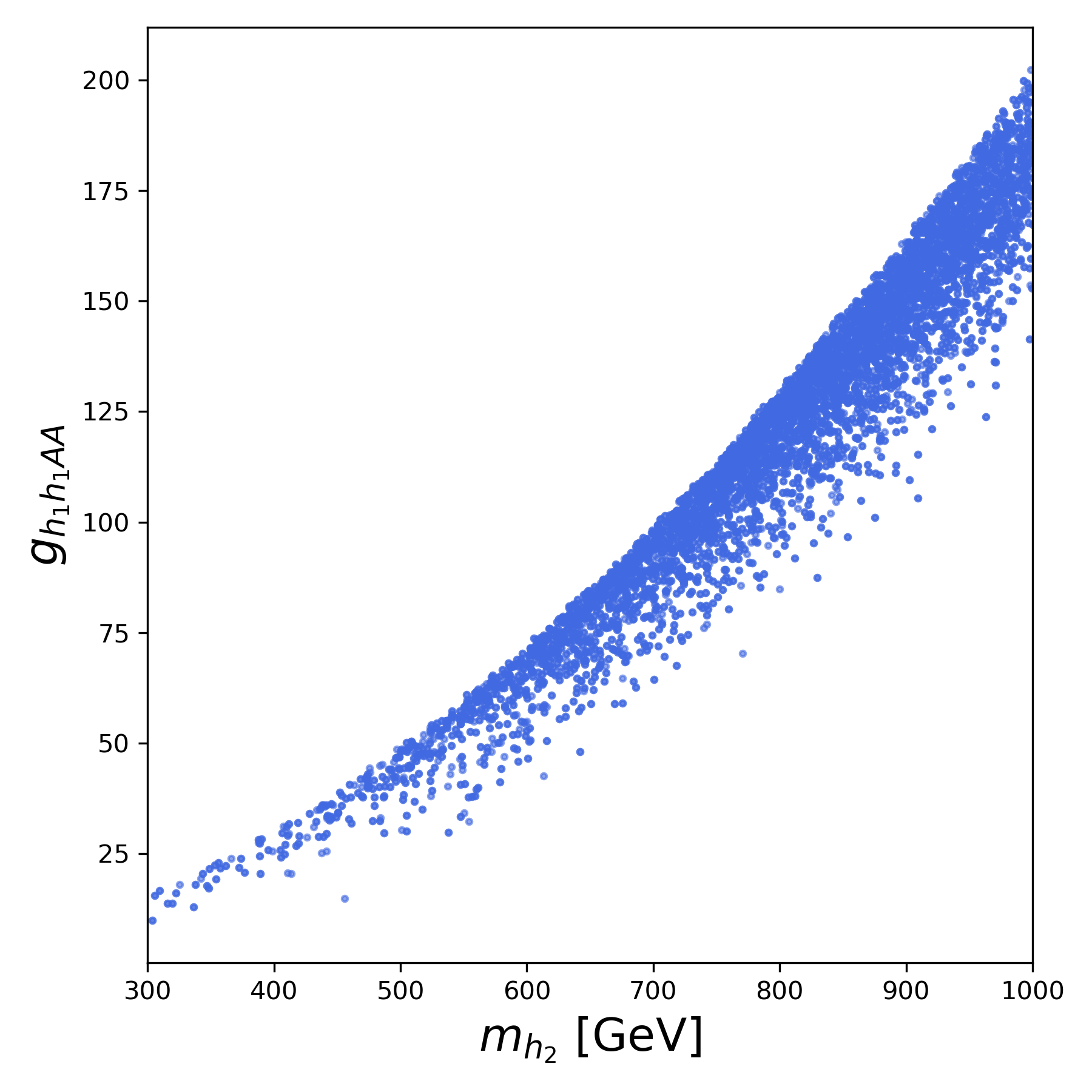}

      \caption{\label{fig::g11AA} Distribution of $g_{h_1 h_1 AA}$ after requirement of SFOEWPT, DM constraints and heavy Higgs searches at LHC. The magnitude of $g_{h_1 h_1AA}$ increases as $m_{h_2}$ increases.}
    \end{figure}

This $a_1$-$m_{h_2}$ correlation leads to an increase in the cross section of certain processes in CSPs. Specifically, the process $pp\to h_1^*\to h_{1}AA$ with diagram-\ref{fig::feyn_diag}(b) is found to be reinforced and even becomes the dominant process for heavy $h_2$ masses. Its cross section is proportional to $g_{h_1h_1AA}$, which can be expressed as
     \begin{align}\nonumber
         g_{h_1 h_1 AA}&=\frac{1}{4v_0v_s^2}(m_{h_1}^2 v_s^2 \cos^3\theta \sin\theta - m_{h_2}^2 v_s^2 \cos^3\theta \sin\theta \\\nonumber
         &+ \sqrt{2}a_1 v_0 \sin^2\theta + m_{h_2}^2 v_0 v_s \cos^2\theta \sin^2\theta\\
         &+ m^2_{h_1} v_0 v_s \sin^4\theta).
     \end{align}
From the formula, it can be observed that this coupling becomes larger with increasing values of $-a_1$ since the $\sin\theta$ is negative due to the heavy scalar requirement as discussed in Sec.~\ref{sec::EWPT}. The resulting correlation between $m_{h_2}$ and $g_{h_1h_1AA}$ is shown in Fig.~\ref{fig::g11AA}.

\subsection{Analysis and results}\label{subsection::analysis}
In this subsection, we will describe the simulation procedures used to select the signals of b-jets plus MET at the HL-LHC. Monte Carlo samples for both the CSPs signal and background events are generated at a $pp$ collider with a center-of-mass energy of 14 TeV. These samples are then normalized to the integrated luminosity of the HL-LHC, which is set to $3000 fb^{-1}$.

We performed a detailed simulation for the mass points listed in Table~\ref{tab:mass_points}. The remaining three parameters corresponding to each mass point were chosen randomly within the allowed parameter space. However, these parameters are believed to affect the relative contributions of different diagrams in CSPs, thereby impacting the selection efficiency. It is important to note that the exclusion reach in the $m_{h_2}-m_A$ plane obtained from our search is intended to be general. Hence, variables that could potentially provide discrimination power between the most dominant diagrams, such as the angular separation of the $b\bar{b}$ system and missing transverse momentum, were not considered in this analysis.

The signal Monte Carlo (MC) samples are generated at the leading order using MadGraph5\_aMC@NLO~\cite{Alwall:2011uj} with UFO and parameter relationships implemented by the FeynRules~\cite{Alloul:2013bka}. The events are then processed through Pythia8~\cite{Sjostrand:2007gs} for parton showering and hadronization. Finally, the simulated events are passed through Delphes3~\cite{deFavereau:2013fsa} to account for the detector response.

\begin{table}[!htbp]
	\centering
	\begin{tabular}{|c|c|c|c|}	\hline
		& ~Process~ & ~$\sigma$ (pb)~ & ~Generator~  \\ \hline
		~ttbar~ & ~$t\bar{t}$~ & 493 & ~Pythia8~ \\ \hline
		~single-top~ & ~$tq$~ & 172 & ~Pythia8~ \\ \hline
		\multirow{2}{*}{~Vh~} & ~$Wh$~ & 0.227 & ~Pythia8~ \\ \cline{2-4}
		& ~$Zh$~ & 0.0768 & ~Pythia8~\\ \hline
		\multirow{2}{*}{~diboson~} & ~$WZ$~ & 4.94 & ~Pythia8~\\ \cline{2-4}
		& ~$ZZ$~ & 1.25 & ~Pythia8~ \\ \hline
		\multirow{2}{*}{~V+jets~} & ~$W+jets$~ & 55.8 & ~MG5\_aMC~ \\ \cline{2-4}
		& ~$Z+jets$~ & 218 & ~MG5\_aMC~ \\ \hline
	\end{tabular}
	\caption{\label{table::bkg_xsec} Information of the background MC samples. The cross sections ($\sigma$) are calculated with the requirement that there are at most one lepton, at least one neutrino and at least one bottom quark in the final states.}
\end{table}

	Associated background processes from top quark pair production (ttbar), single top quark production (single-top), Vh production, diboson production, and processes involving a vector boson in association with jets (V+jets) are generated using Pythia8~\cite{Sjostrand:2007gs}. The aim is to simulate backgrounds that have similar visible final states as our target signal and can contaminate into the signal region. Therefore, all background events are required to have at most one lepton and at least one bottom quark. Additionally, they must have at least one neutrino to satisfy the requirement of high missing transverse energy. Table~\ref{table::bkg_xsec} provides a summary of the background generation process. The showering and simulation approach for background events follows the same procedure as for the signal.

	The generated Monte Carlo samples are analyzed using MadAnalysis5~\cite{Conte:2012fm}. During the object reconstruction stage, some basic requirements on transverse momentum and pseudorapidity are applied. Specifically, jets are required to have $p_t > 25~\text{GeV}$ and $|\eta| < 2.5$, while electrons and muons are required to have $p_t > 10~\text{GeV}$ and $|\eta| < 2.4$. These requirements help ensure the quality and reliability of the reconstructed objects in the analysis.

	\begin{figure}[!hbp]\centering 
		\begin{center}
			\includegraphics[width=0.49\textwidth]{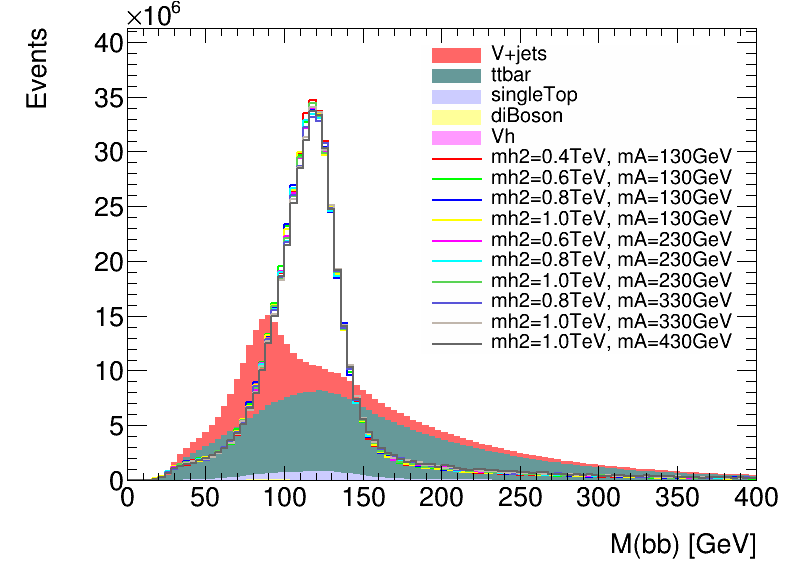}
		\end{center}
		\caption{\label{fig::m_bb} Distributions of the invariant mass of bottom-pair system after the first two cuts.}
	\end{figure}
	\begin{figure}[!hbp]\centering 
		\begin{center}
			\includegraphics[width=0.49\textwidth]{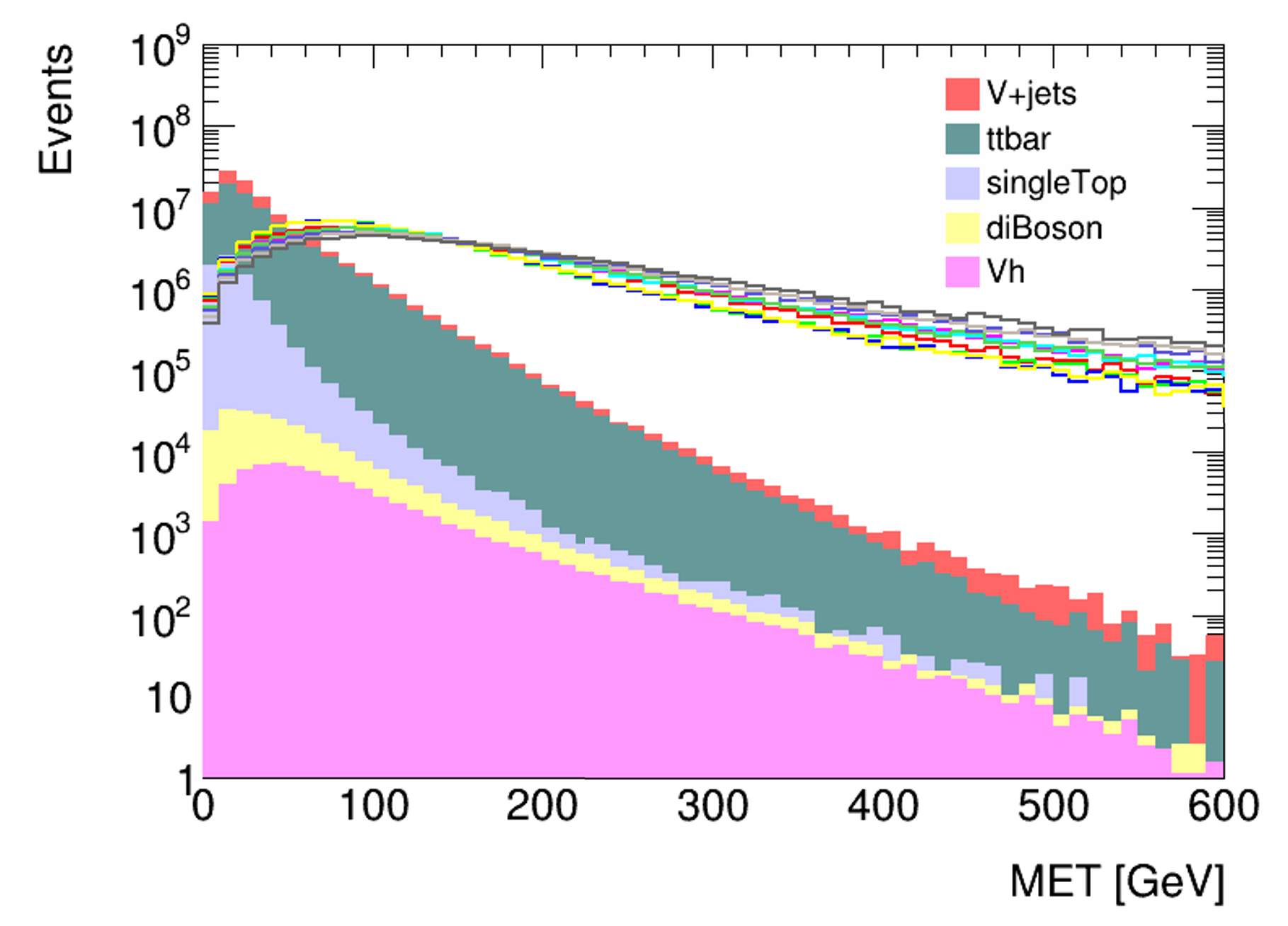}
		\end{center}
		\caption{\label{fig::MET} Distributions of the missing transverse energy after the first three cuts.}
	\end{figure}
	
    Two general cuts are initially applied to distinguish the signal and background events for all mass points:
 \begin{itemize}
     \item Cut-1 $n_{\rm{lepton}}=0$.
     \item Cut-2 $n_{b-\rm{jets}}=2$. 
 \end{itemize}

 After applying these cuts, we present the distribution of the invariant mass of the $b\bar{b}$ system in Fig.~\ref{fig::m_bb}. In this figure, the signal events have been rescaled to match the remaining background events. To identify the bottom-quark pair from the SM-like Higgs boson decay, we implement a related cut:
 \begin{itemize}
\item Cut-3: $100~\mathrm{GeV} < m_{b\bar{b}} < 140~\mathrm{GeV}$.
\end{itemize}
Furthermore, we take into account the missing transverse energy to further distinguish signal events from the background. As depicted in Fig.~\ref{fig::MET}, this variable is expected to be significantly large in our signal samples. To ensure that the statistical uncertainty of the generated background does not have a substantial impact, we apply a relatively loose cut:
\begin{itemize}
\item Cut-4: $\mathrm{MET} > 550~\mathrm{GeV}$
\end{itemize}
The purpose of this cut is to ensure that the signal events can be effectively separated from the background.
     
After applying the selection criteria, the number of the signal events that can be detected at a 95\% confidence level corresponds to a cross section close to $10^{-2}~\rm{pb}$. The exact exclusion cross sections are listed in Table~\ref{tab:exclusion}. From the table, we observe that the selection efficiency is primarily dependent on the mass of the dark matter candidate $A$, as one can expect from the MET cut.  To cover the entire $300~\rm{GeV} < m_{h_2} < 1000~\rm{GeV}$ range for each $m_A$ point, we employ the linear interpolation and extrapolation based on the limits obtained from our analysis. In particular for the case of $m_A=430$ GeV, we make the assumption that the limit remains constant throughout the entire range. Notice that the expected discovery ability is enhanced in $m_{h_2}=400$ GeV for the case of $m_A=130$ GeV, therefore, our approach to obtain the upper limit in the low $m_{h_2}$ region can be considered conservative, the actual exclusion limit in that region might be even stronger than what is indicated by our study. Subsequently, we employ a bivariate spline approximation based on this rectangular mesh to obtain a wide range of upper limits on the $m_{h_2}-m_A$ plane.
 
\begin{table}[htbp]
  \centering
  \resizebox{\textwidth/2}{!}{
    \begin{tabular}{| c | c | c | c | c |}\hline
      & $m_A=130$ GeV & $m_A=230$ GeV & $m_A=330$ GeV & $m_A=430$ GeV\\ \hline
     $m_{h_2}=400$ GeV & 7.9 fb &  &  & \\ \hline
     $m_{h_2}=600$ GeV & 9.5 fb & 4.4 fb &  & \\ \hline
     $m_{h_2}=800$ GeV & 9.7 fb & 4.6 fb & 3.1 fb & \\ \hline
     $m_{h_2}=1000$ GeV & 9.3 fb & 4.7 fb & 2.9 fb & 2.2 fb\\ \hline
     \end{tabular}
}
  \caption{The exclusion cross sections at a 95\% confidence level for each mass point in the analysis.}
  \label{tab:exclusion}
\end{table}

\begin{figure}[!tbp]\centering 

        \includegraphics[width=0.49\textwidth]{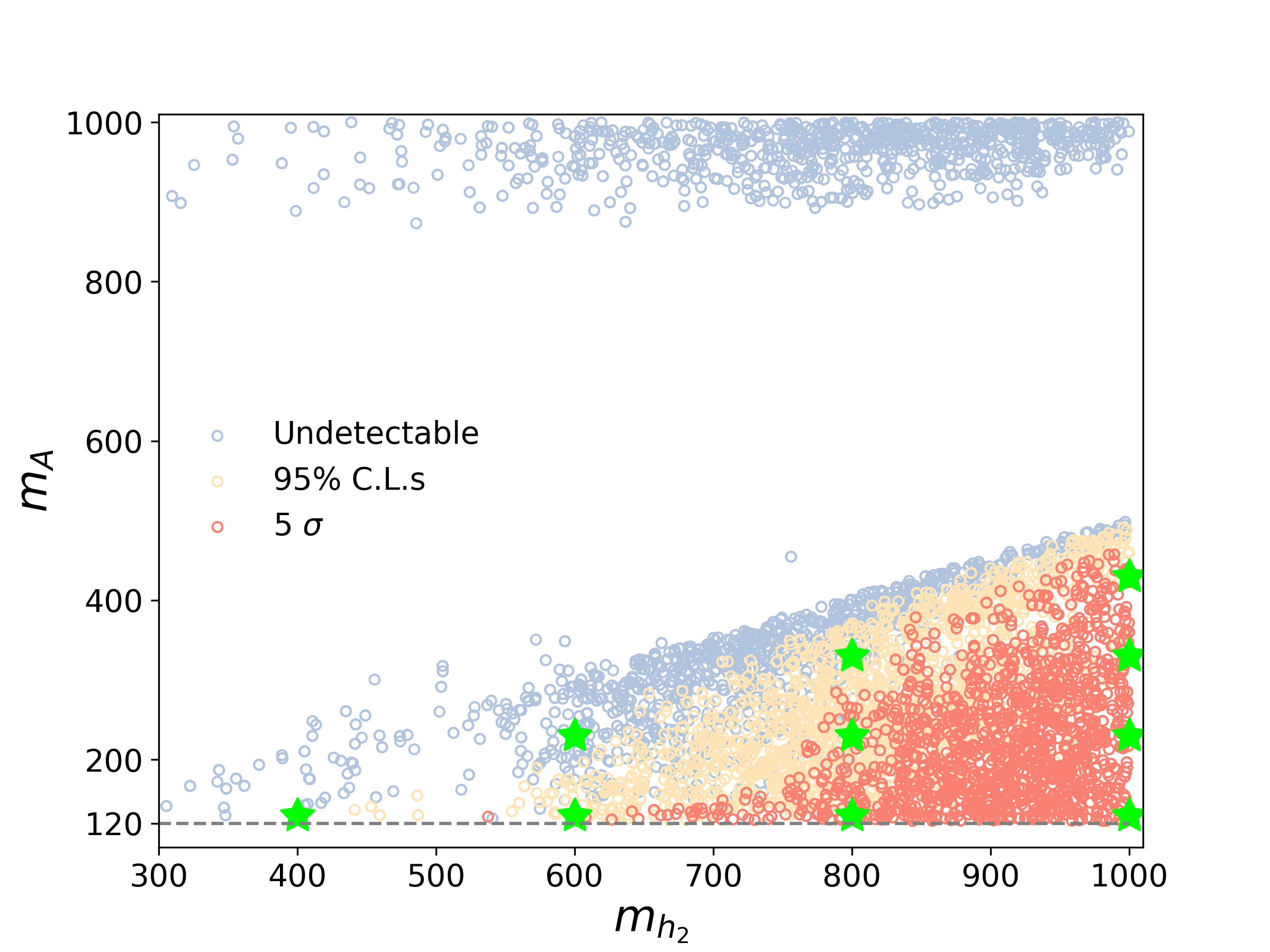}
        \includegraphics[width=0.49\textwidth]{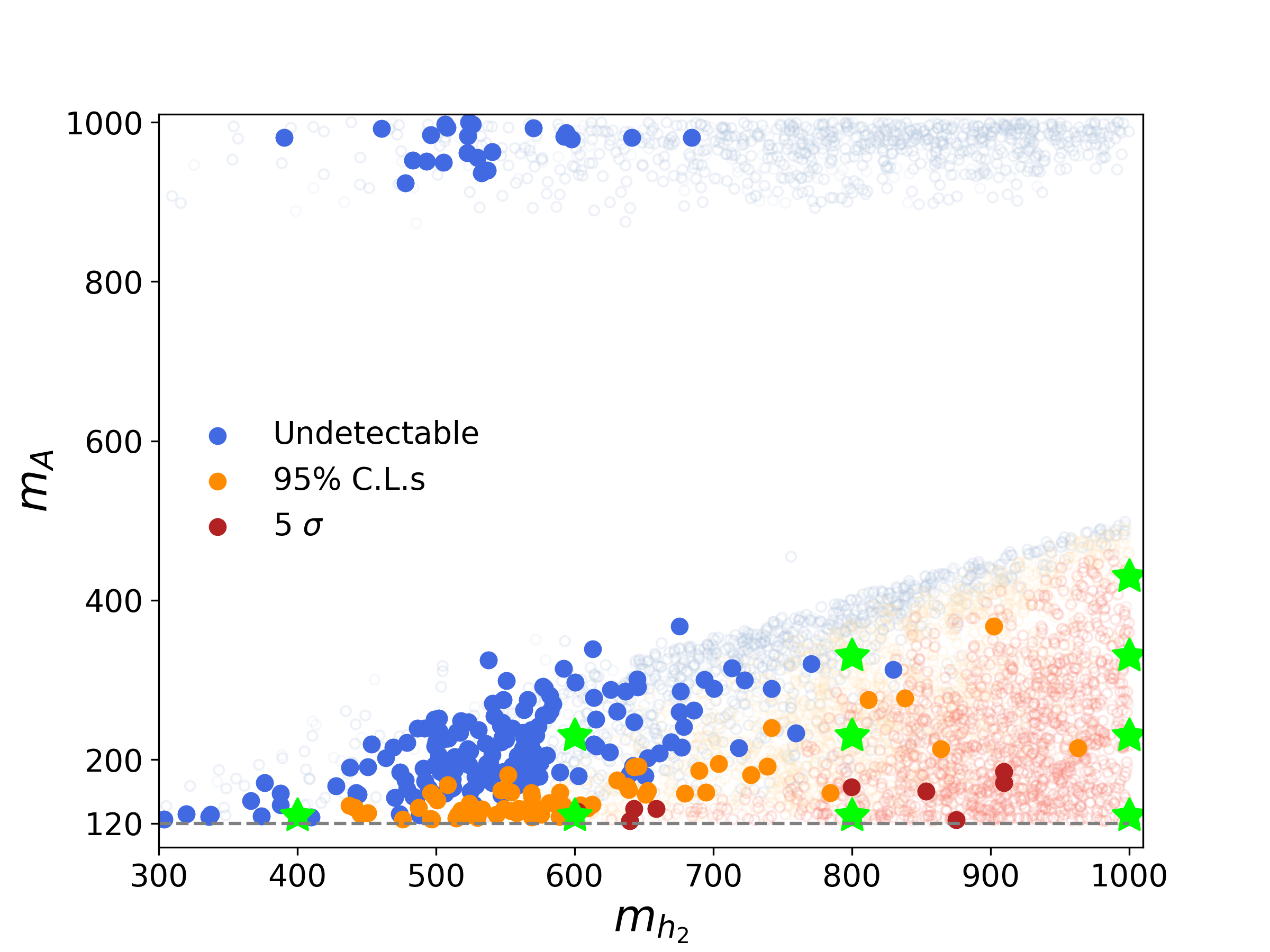}

      \caption{\label{fig::remain}  Exclusion/Discovery plot on $m_{h_2}-m_A$ plane.  Upper panel (light-colored points): Space that can be probed by both $b\Bar{b}$+MET and regular resonace search. Lower panel (dark-colored points): Space that cannot be probed by regular resonace searches but is able to be probed by $b\Bar{b}$+MET channel. The meaning of the color is shown in the plot. Green stars are the benchmarks listed in the Table~\ref{tab:exclusion}.}
\end{figure}

 We then scatter the parameter points from our general scanning space (Eq.\ref{eq::scan_space}) on the this two-dimensional plane, taking into account all the current experimental constraints. The resulting plot is shown in Fig.~\ref{fig::remain}. Green stars are the benchmark mass points that are used in the analysis. All points generate a SFOEWPT and satisfy DM relic density as well as direct detection constraints. The light colored points are expected to be excluded by the HL-LHC, while the dark ones are not. The dark red and dark orange points are expected to be probed by the $b\bar{b}$+MET channel beyond $5\sigma$ and $1.96\sigma$ , respectively. The dark blue points survive all the searches and remain after selection. 
 
Several observations are in order:
   
Firstly, in Fig.~\ref{fig::remain}, there is a distinct line with a positive slope at the upper boundary of the mass points region, particularly noticeable for heavier $h_2$ values. This slope corresponds to the relationship $m_{h_2}=2m_A$. The region located above this slope is largely excluded based on the results of the heavy scalar resonance search, as discussed in Section~\ref{sec::bound_lhc}.

Secondly, the density of red points is more pronounced in the region of heavier $h_2$ masses, suggesting a more promising discovery potential in the higher $h_2$ mass range. At first glance, this result may seem counter-intuitive. However, it can be attributed to the increasing cross section of the $pp\to h_1^*\to h_1AA$ process, as discussed in Subsection~\ref{subsection::CSPs}.

Finally, a significant portion of the parameter space with heavier $h_2$ masses can be effectively probed by the $b\bar{b}$+MET search at the HL-LHC.
As the $b\bar{b}$+MET signal distinguishes the cxSM from the xSM, its use could allow one to discover the SFOEWPT-viable cxSM for heavy Higgs Mass $\gtrsim 650$ GeV at a $5\sigma$ confidence level. Moreover, such a search for this channel would complement other resonance searches for $m_{h_2}\gtrsim 450~\text{GeV}$, since a considerable number of parameter points, which are likely to remain viable after these resonance searches, can be effectively probed through the proposed $b\bar{b}$+MET channel.
		
    	\section{Conclusion}\label{sec::conclusion}

	Through spontaneous and soft breaking of a global U(1) symmetry, the cxSM introduces two additional degrees of freedom, with one catalyzing a possible SFOEWPT and the other providing a viable DM candidate.
	Previous studies have demonstrated the viability of the cxSM for both DM and SFOEWPT and have elucidated the correlation between the singlet scalar-SM Higgs coupling and the occurrence of a SFOEWPT in the cxSM parameter space. In addition, there exists a coupling $h_1 A A$ between the SM-like Higgs and pseudscalar (DM) pair. For sufficiently light $A$, the Higgs invisible decay is induced for small pseudoscalar masses.
	To avoid an experimentally excluded excess of the Higgs invisible decay, one way is to restrict $m_A$ to a narrow window around $m_{h_2}/2$ or to implement a delicate fine tuning cancellation between $a_1$ and $v_s$. Alternatively, one may take $m_A>m_{h_2}/2$ so that the Higgs invisible decay is impossible.
	In both cases, a  distinctive signal in $pp$ collisions is a $b{\bar b}$ pair plus MET, with various contributions being mediated by on- and/or off-shell $h_{1,2}$ bosons.
	Searches for such signal processes have never been performed for the cxSM.
	Therefore, there exists strong motivation to study the HL-LHC reach for the cxSM EWPT-DM viable parameter space.      
	
	In this work, we have performed a detailed analysis of this reach. 
	The previously considered most relevant heavy resonance searches at the LHC, which include the di-Higgs channels (see in Fig.~\ref{fig::scalar_ATLAS_d}) and the $WW + ZZ$ channels (see in Fig.~\ref{fig::scalar_ATLAS_a} and Fig.~\ref{fig::scalar_ATLAS_b}), are powerful in probing $m_{h_2} \gtrsim 700$ GeV and $m_{h_2} \lesssim 450$ GeV in HL-LHC respectively.
    Thus, viable EWPT and DM parameters are hopeful to be observed/excluded in these two areas. 
    Note that in the second area, a key EWPT related parameter, the mixing angle $\sin\theta$, that is proportional to $\delta_2 v_s$ is constrained by the di-boson channel.
    On the contrary, outside this area, the optional parameter space of $\sin\theta$ is larger (see in Fig.~\ref{fig::correlation_sin}). 
    As a complement to the resonance searches and a key feature signal of the cxSM, the $b\bar{b}+\text{MET}$ search  provides a 
    feasible means for probing a considerable portion of the viable parameter space where $m_{h_{2}} \gtrsim 450$ GeV and for distinguishing the cxSM from the xSM for $m_{h_{2}} \gtrsim 650$ GeV.

    While we considered a complete set of processes with $b\Bar{b}+\text{MET}$ final states, we designed the detection method based on the characteristics of the heavy scalar resonance signal events.
    We find that one of the dominant processes, $p p \to h^{*}_1 \to h_1 A A$ that is induced by the coupling $g_{h_1 h_1 A A}$, is reinforced significantly by the increasing $-a_1$ in heavy $m_{h_2}$ region.

\section{Acknowledgement}

M.J. Ramsey-Musolf and W. Zhang were supported in part by the National Natural Science Foundation of China under grant no. 11975150 and by the Ministry of Science and Technology of China under grant no. WQ20183100522. M. J. Ramsey-Musolf also gratefully acknowledges support under the Double First Class Plan of the Shanghai Jiao Tong University and sponsorship from Shanghai Tang Junyuan Education Foundation. Y. Cai received financial support from the China Scholarships Council program. L. Zhang's work was supported by the National Science Fund of China for Excellent Young Scholars under grant number 12122507.

    \appendix

    \section{Oblique Parameter}\label{sec::appendix_EWPO}

Following the notation by Peskin and Takeuchi~\cite{Peskin:1991sw}, the contribution to S, T and U from the new scalar can be expressed as~\cite{Degrassi:1993kn, Chen:2008jg, Dawson:2009yx, Profumo:2014opa}

{\footnotesize \begin{align*}
&\begin{aligned}
    \Delta S &= \frac{1}{\pi}|\sin\theta|^2\{B_0(0,m_{h_2},M_Z)-B_0(M_Z,m_{h_2},M_Z) \\
             &~~~+ \frac{1}{M_Z^2} \left[B_{22}(M_Z,m_{h_2},M_Z)-B_{22}(0,m_{h_2},M_Z)\right] \},
\end{aligned} \\
&\begin{aligned}
    \Delta T &= \frac{1}{4\pi s_w^2}|\sin\theta|^2\{ -B_0(0,m_{h_2},M_W)+\frac{1}{c_w^2}B_0(0,m_{h_2},M_Z) \\
             &~~~+ \frac{1}{M_W^2} \left[B_{22}(0,m_{h_2},M_W)-B_{22}(0,m_{h_2},M_Z)\right] \},
\end{aligned} \\
&\begin{aligned}
    \Delta (U+S) &= \frac{1}{\pi}|\sin\theta|^2\{ B_0(0,m_{h_2},M_W)-B_0(M_W,m_{h_2},M_W) \\
                 &~~~+ \frac{1}{M_W^2} \left[-B_{22}(0,m_{h_2},M_W)+B_{22}(M_W,m_{h_2},M_W)\right] \},
\end{aligned}
\end{align*}}
where $B_0$ and $B_{22}$ are Passarino-Veltman funtions~\cite{Passarino:1978jh}. 
    
    \section{Three-body decay phase space}\label{sec::appdix_3_body}

We use the example of the "three-body" decay process, where the differential cross section for a process of a three-body decay is
\begin{equation}
	d\Gamma=\frac{(2\pi)^4}{2m_{h_2}}|\mathcal{M}|^2 d\Phi_3,
\end{equation}
where we use the $d\Phi_n$ to denote the n-body phase space.
Since the standard form of the phase space volume element with n final state particles can be decomposed into a number of multiplication of 2-body phase space, with the $d\Phi_3$ is related with  according to
\begin{align}
	d\Phi_3 &= d\Phi_2(m_{AA},m_{A},m_{A}) d\Phi_2(m_{h_2},m_{AA},m_{h_1})(2\pi)^2 dm_{AA}^2 \\ \nonumber
	&= d\Omega^* \frac{|p^*|}{(2\pi)^6 4m_{AA}} d\Omega_3 \frac{|p_3|}{(2\pi)^6 4M} (2\pi)^2 
	dm_{AA}^2,
\end{align}
where the $\Omega^*$ and $\Omega_3$ are the solid angles of the off-shell SM-like Higgs and heavy resanance respectively.

The integration parameter, $m_{AA}$, is the invariant mass of two-DM system. The integration range is $\left[2m_A, m_{h_2}-m_{h_1}\right]$. $p^*$ ($p_3$) is the momentum of off-shell (on-shell) SM-like Higgs momentum.
Thus the differential cross section can be expressed as 

\begin{align}\label{eq::h2_xsec}\nonumber
	d\Gamma &= \frac{(2\pi)^5}{16M^2}|\mathcal{M}|^2 dm_{AA}
	d\Omega^* d\Omega_3 \\ \nonumber
	&= \frac{\lambda^{\frac{1}{2}}(m_{AA}, m_A, m_A) \lambda^{\frac{1}{2}}(m_{h_2},m_{AA},m_{h_1})}{32\pi^3 m_{h_2}^2}\\ 
	&~~~\times |\frac{g_{211}g_{1AA}}{m_{AA}^2}|^2 ~dm_{AA}.
\end{align}
where we have used $|\mathcal{M}|^2 = |\frac{g_{211}g_{1AA}}{m_{AA}^2}|^2$ and

\begin{equation}\label{eq::lambda}
	\lambda^{\frac{1}{2}}(m_{12}, m_1, m_2) = \frac{\sqrt{\left[m_{12}^2-\left(m_1^2+m_2^2\right)\right]^2-4m_1^2m_2^2}}{2m_{12}}.
\end{equation}

Based on these relations, we calculate both "2-body" and "3-body" branching ratios and scan over the general parameter space via

\begin{equation}
	BR(h_2 \to h_1 A A)=\frac{\Gamma_{h_2 \to h_1 A A}}{\sin^2{\theta} ~\Gamma^{SM}_{h} + \Gamma_{h_2 \to A A} + \Gamma_{h_2 \to h_1 A A}}
\end{equation}
for "3-body" case, and the "2-body" case has a similar form.
    
    \section{Additional Content for HL-LHC search}\label{sec::appendix_lhc}

\begin{figure*}[htbp]\centering 
    \begin{center}
        \centering
	\includegraphics[width=1\textwidth]{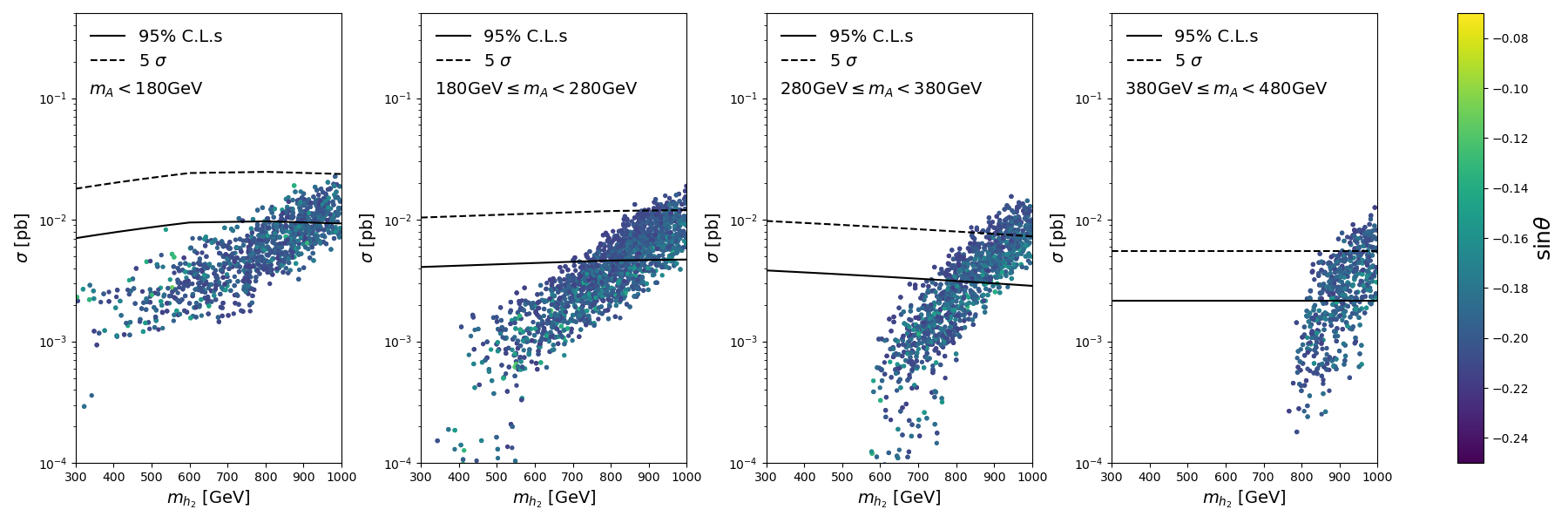}
    \end{center}
    \caption{\label{fig::xsec} Cross section distributions and detectable limits for CSPs. The points plotted satisfy all the constraints discussed in the previous sections. From left to right, the sub-figures correspond to the parameter points with $m_A$ around 130 GeV, 230 GeV, 330 GeV, and 430 GeV, respectively. The dashed line and the solid line in each sub-figure represent the $5~\sigma$ and $1.96~\sigma$ discovery significance in the HL-LHC via our analysis for the corresponding $m_A$ values of 130 GeV, 230 GeV, 330 GeV, and 430 GeV. The color-bar representing value of $\sin\theta$.}
\end{figure*}

The cross sections of parameter points surviving all of the current experimental constraints are shown in Fig.~\ref{fig::xsec} for $m_A$ around 130GeV, 230GeV, 330GeV and 430GeV respectively. {Notice that the cross section increases as $m_{h_2}$ increases. The reason is that the coupling $g_{11AA}$ grows with $m_{h_2}$. Hence the cross section of the dominant process $pp\to h_1^* \to h_1 A A$ increases.} The dashed line and the solid line in each sub-figure represent the $5~\sigma$ and $1.96~\sigma$ discovery significance in the HL-LHC via our analysis for the corresponding $m_A$ values of 130 GeV, 230 GeV, 330 GeV, and 430 GeV.

    \section{1-loop RGE running for cxSM}\label{sec::appdix_RGE}
We derive the renormalization group equations (RGEs) by using the python package \texttt{PyR@TE 3} in the $\overline{{\rm MS}}$ scheme at the 1-loop level.

\begin{align}
\beta^{(1)}(g_1) &= \frac{41}{10} g_1^{3} \\
\beta^{(1)}(g_2) &= -\frac{19}{6} g_2^{3} \\
\beta^{(1)}(g_3) &= -7 g_3^{3} \\
\beta^{(1)}(d_2) &= 2\delta_2^{2} + 5 d_2^{2} \\
\nonumber
\beta^{(1)}(\lambda) &= 6 \lambda^{2} + \delta_2^{2} - 3 g_1^{2} \lambda - 9 g_2^{2} \lambda + \frac{3}{2} g_1^{4} + 3 g_1^{2} g_2^{2} + \frac{9}{2} g_2^{4} \\ 
&+ 12 \lambda \abs{Y_t}^2 - 24 \abs{Y_t}^4 \\
\beta^{(1)}(\delta_2) &= 3 \delta_2 \lambda + 2 \delta_2^{2} + 2 d_2 \delta_2 - \frac{3}{2} \delta_2 g_1^{2} - \frac{9}{2} \delta_2 g_2^{2} + 6 \delta_2 \abs{Y_t}^2 \\
\beta^{(1)}(d_2) &= 2 \delta_2^{2} + 5 d_2^{2} \\
\beta^{(1)}(Y_t) &= \frac{9}{2} Y_t\abs{Y_t}^2 - \frac{17}{12} g_1^{2} Y_t - \frac{9}{4} g_2^{2} Y_t - 8 g_3^{2} Y_t,
\end{align}
where the couplings at scale $\mu = 246$ GeV are
\begin{align}\nonumber
    &g_1=\sqrt{\frac{5}{3}}\times 0.36001, ~~g_2=0.64632, ~~g_3=1.15330 \\
    &y_t=0.93930, ~~y_b=0.01897.
\end{align}

    \newpage
	\vspace{0.5cm}	
	\phantomsection
	\addcontentsline{toc}{section}{References}
	\bibliography{cxSM}
	\bibliographystyle{utphys}
\end{document}